\documentclass[twocolumn,showpacs,amsmath,amssymb,aps,prx,footinbib,floatfix,superscriptaddress,longbibliography]{revtex4-1}
\usepackage{graphicx}
\usepackage{braket}
\usepackage{savesym}
\usepackage{amsfonts}
\usepackage{bm}
\usepackage{color}
\usepackage{subfigure}
\usepackage{wasysym}
\usepackage{upgreek}
\usepackage{float}
\usepackage{dsfont}
\usepackage{mathtools}
\usepackage{multirow}
\usepackage[usenames,dvipsnames,svgnames,table]{xcolor}
\usepackage{hyperref}
\usepackage{hhline} 
\usepackage{multirow} 
\usepackage[normalem]{ulem}

\usepackage{makecell}
\usepackage[english]{babel}

\hypersetup{
	colorlinks=true,       
	linkcolor=VioletRed,  
    citecolor=JungleGreen,        
    filecolor=Orchid,      
    urlcolor=black           
}
\begin{document}

    \title{Quantum HyperNetworks: Training Binary Neural Networks in Quantum Superposition}

\author{Juan Carrasquilla}
\email{jcarrasquill@ethz.ch}
\affiliation{Institute for Theoretical Physics, ETH Zürich, 8093, Switzerland}
\affiliation{Vector Institute, MaRS  Centre,  Toronto,  Ontario,  M5G  1M1,  Canada}
\affiliation{Department of Physics and Astronomy, University of Waterloo, Waterloo, Ontario, N2L 3G1, Canada}

\author{Mohamed Hibat-Allah}
 \affiliation{Perimeter Institute for Theoretical Physics, Waterloo, ON N2L 2Y5, Canada}
\affiliation{Vector Institute, MaRS  Centre,  Toronto,  Ontario,  M5G  1M1,  Canada}
\affiliation{Department of Physics and Astronomy, University of Waterloo, Waterloo, Ontario, N2L 3G1, Canada}

\author{Estelle Inack}
 \affiliation{Perimeter Institute for Theoretical Physics, Waterloo, ON N2L 2Y5, Canada}
 \affiliation{yiyaniQ, Toronto,  Ontario,  M4V 0A3,  Canada}
\affiliation{Department of Physics and Astronomy, University of Waterloo, Waterloo, Ontario, N2L 3G1, Canada}

\author{Alireza Makhzani}
\affiliation{Vector Institute, MaRS  Centre,  Toronto,  Ontario,  M5G  1M1,  Canada}
\affiliation{University of Toronto, Toronto, Ontario M5S 1A7, Canada}

\author{Kirill Neklyudov}
\affiliation{Vector Institute, MaRS  Centre,  Toronto,  Ontario,  M5G  1M1,  Canada}

\author{Graham W. Taylor}
\affiliation{School of Engineering, University of Guelph, Guelph, Ontario, ON N1G 2W1, Canada}
\affiliation{Vector Institute, MaRS  Centre,  Toronto,  Ontario,  M5G  1M1,  Canada}

\author{Giacomo Torlai}
\thanks{Work done before joining AWS.}
\affiliation{AWS Center for Quantum Computing, Pasadena, CA, USA }

\date{\today}

\begin{abstract}
Binary neural networks, i.e., neural networks whose parameters and activations are constrained to only two possible values,  offer a compelling avenue for the deployment of deep learning models on energy- and memory-limited devices. However, their training, architectural design, and hyperparameter tuning remain challenging as these involve multiple computationally expensive combinatorial optimization problems. Here we introduce quantum hypernetworks as a mechanism to train binary neural networks on quantum computers, which unify the search over parameters, hyperparameters, and architectures in a single optimization loop. Through classical simulations, we demonstrate that our approach effectively finds optimal parameters, hyperparameters and architectural choices with high probability on classification problems including a two-dimensional Gaussian dataset and a scaled-down version of the MNIST handwritten digits. We represent our quantum hypernetworks as variational quantum circuits, and find that an optimal circuit depth maximizes the probability of finding performant binary neural networks. Our unified approach provides an immense scope for other applications in the field of machine learning.

\end{abstract}

\maketitle

\section{Introduction}

The availability of high quality data sources along with algorithmic and hardware advances for training neural networks have paved the way for a new generation of large models displaying unprecedented accuracy across a wide array of technologically and scientifically relevant tasks. These advances crucially depend on the availability of specialized computational resources such as graphics and tensor processing units, which demand a high electricity consumption. In particular, a set of key but computationally expensive elements in the modern machine learning (ML) workflow include hyperparameter optimization and neural architecture search. Traditionally, these operate via an outer optimization loop which searches through the hyperparameter and architectural state spaces guided by the model's performance on a validation set, and an inner optimization which adjusts the parameter of the neural network on a training set. Such a nested optimization process remains the most computationally demanding task in the modern ML workflow and entails an unsustainable carbon footprint, which calls for computationally efficient hardware and algorithms to train and search for neural architectures~\cite{strubellEnergyPolicyConsiderations2019}.

Neural networks with binary parameters and activations (BiNNs) partially alleviate these issues as they are computationally efficient, hardware-friendly, and energy efficient. Beyond a direct 32-fold reduction of the memory footprint with respect to a full-precision neural network, BiNNs can exploit specialized hardware implementations that simultaneously increase computational speed~\cite{rastegariXNORNetImageNetClassification2016} and improve their energy efficiency~\cite{DBLP:journals/corr/HanMD15}. Another benefit of very low-precision neural networks is their improved robustness against adversarial attacks while matching the performance of full-precision models in the worst cases~\cite{galloway2018attacking}. While in principle it is possible to binarize trained continuous-variable neural networks, such a procedure typically leads to significant accuracy losses, which makes it preferable to directly learn their binary parameters. 

The ML community has developed approaches to the use of BiNNs which bypass the infeasible discrete optimization of their training through a re-framing of the problem in the conventional domain of gradient descent algorithms. These include post-quantization of conventionally trained neural networks, as well as deterministic and stochastic relaxations of the original problem both for parameter tuning
\citep{han2015deep,hubara2016binarized,ullrich2017soft,pmlr-v119-meng20a} and architecture search~\cite{bulatBATSBinaryArchitecTure2020}. In spite of these advances, the combined optimization of a BiNN's parameters and their associated hyperparameter and architecture searches remain computationally demanding as these involve solving multiple nested combinatorial optimization problems or their associated relaxations.


Quantum computing utilizes quantum interference and entanglement to tackle computationally challenging problems, offering an alternative for training neural  networks~\cite{nevenTrainingBinaryClassifier2008,silvaSuperpositionBasedLearning2010,Quanta65,Verdon2017,baldassiEfficiencyQuantumVs2018,Verdon2018,dos_Santos2018,allcockQuantumAlgorithmsFeedforward2020,liaoQuantumSpeedupGlobal2021,zlokapaQuantumAlgorithmTraining2021,tortaQuantumApproximateOptimization2021,alarconAcceleratingTrainingSinglelayer2022,nikoloskaQuantumAidedMetaLearningBayesian2022,lamiQuantumAnnealingNeural2022}.  Notably, quantum annealing, as explored in Ref.~\cite{baldassiEfficiencyQuantumVs2018}, demonstrated an exponential speed-up compared to classical simulated annealing for a binary perceptron problem, a theoretical model of classification task for a single-layer neural network.  

These speedups arise because the energy landscape of neural network cost functions encompasses numerous suboptimal metastable states and regions with densely packed ground states~\cite{baldassiEfficiencyQuantumVs2018}. In machine learning, dense low-energy configurations play a critical role in model generalization by providing resilience against fluctuations in weight configurations, reducing susceptibility to overfitting. While training with classical simulated annealing tends to get trapped in the metastable states, Ref.~\cite{baldassiEfficiencyQuantumVs2018} revealed that quantum annealing helps navigate these dense ground state regions efficiently. Given the close connection between quantum annealing and algorithms like the quantum approximate optimization algorithm~\cite{farhiQuantumApproximateOptimization2014,bradyOptimalProtocolsQuantum2021a}, it stands to reason that quantum algorithms may excel in finding parameter models within dense regions with low generalization error. 

Additionally, the training of binary neural networks can be understood as a blackbox binary optimization problem for which successful variational quantum algorithms showcase competitive performance compared to classical algorithms~\cite{zoufalVariationalQuantumAlgorithm2023} as well as displayed improvements over quantum annealing for binary perceptrons~\cite{tortaQuantumApproximateOptimization2021}. Therefore, we focus on variational quantum algorithms (VQAs), which have emerged as promising approaches for achieving quantum computational advantage on near-term quantum devices~\cite{Preskill2018quantumcomputing, Cerezo2021}.  These algorithms employ parameterized quantum circuits adaptable to experimental constraints, such as limited qubits, gate infidelities, and errors in realizable quantum circuits ~\cite{Preskill2018quantumcomputing, Cerezo2021}. 

Here we promote hypernetworks--networks that generate the weights of another network~\cite{haHyperNetworks2016}--to quantum hypernetworks, i.e., quantum states that generate the weights of a neural network. Quantum hypernetworks offer an alternative approach to the training of BiNNs through a unification of the parameter, hyperparameter, and architecture searches in a single optimization loop. A quantum hypernetwork, here implemented through  a parameterized quantum circuit of variable depth, is trained to search over an augmented space comprising the parameters of the neural network, its hyperparameters, and any desired architectural choices with an eye on improving the overall efficiency of the BiNN workflow. Through classical simulations, we show that quantum hypernetworks with short depth and limited connectivity can jointly optimize BiNNs' hyperparameters, architectural choices, and parameters for toy classification problems including a two-dimensional Gaussian dataset and a scaled-down version of the MNIST handwritten digits. We find that the probability of finding performant BiNNs is maximized at a specific circuit depth, which suggests that an optimal use of entanglement and quantum effects decrease the probability that the optimization finds poor local minima.

Through a Fourier analysis, we reveal that the objective functions used to train the BiNNs are predominantly local. This observation, together with our numerical experiments, suggests that quantum hypernetworks built from local low-depth circuits with limited connectivity, all of which are common features to most currently available quantum computers, can be effective at training BiNNs. Our analysis indicates that the locality of the objective function may not induce tractability problems related to the presence of barren plateaus which interfere with the accurate estimation of the gradients used during the optimization of the circuits.

\section{Results}

\subsection{Variational Quantum HyperNetworks} 
To encode the problem in a form suitable to optimization by a quantum computer, we consider quantum states composed of $N$ qubits written in the computational basis corresponding to the eigenstates of tensor products of the Pauli operator $\hat{\sigma}^{z}_i$ acting on qubits $i$, namely 
\begin{equation}\label{eq:qstate}
   |\Psi \rangle =  \sum_{\sigma_1,\ldots,\sigma_N} \Psi(\sigma_1,\ldots,\sigma_N) |\sigma_1,\ldots,\sigma_N\rangle, 
\end{equation}
where $\hat{\sigma}^{z}_i|\sigma_1,\ldots,\sigma_i, \ldots,\sigma_N\rangle =(2\sigma_i-1) |\sigma_1,\ldots,\sigma_i, \ldots,\sigma_N\rangle$ , and $\sigma_i \in \{0,1\}$. 
 
\begin{figure}%
    \includegraphics[width =\linewidth]{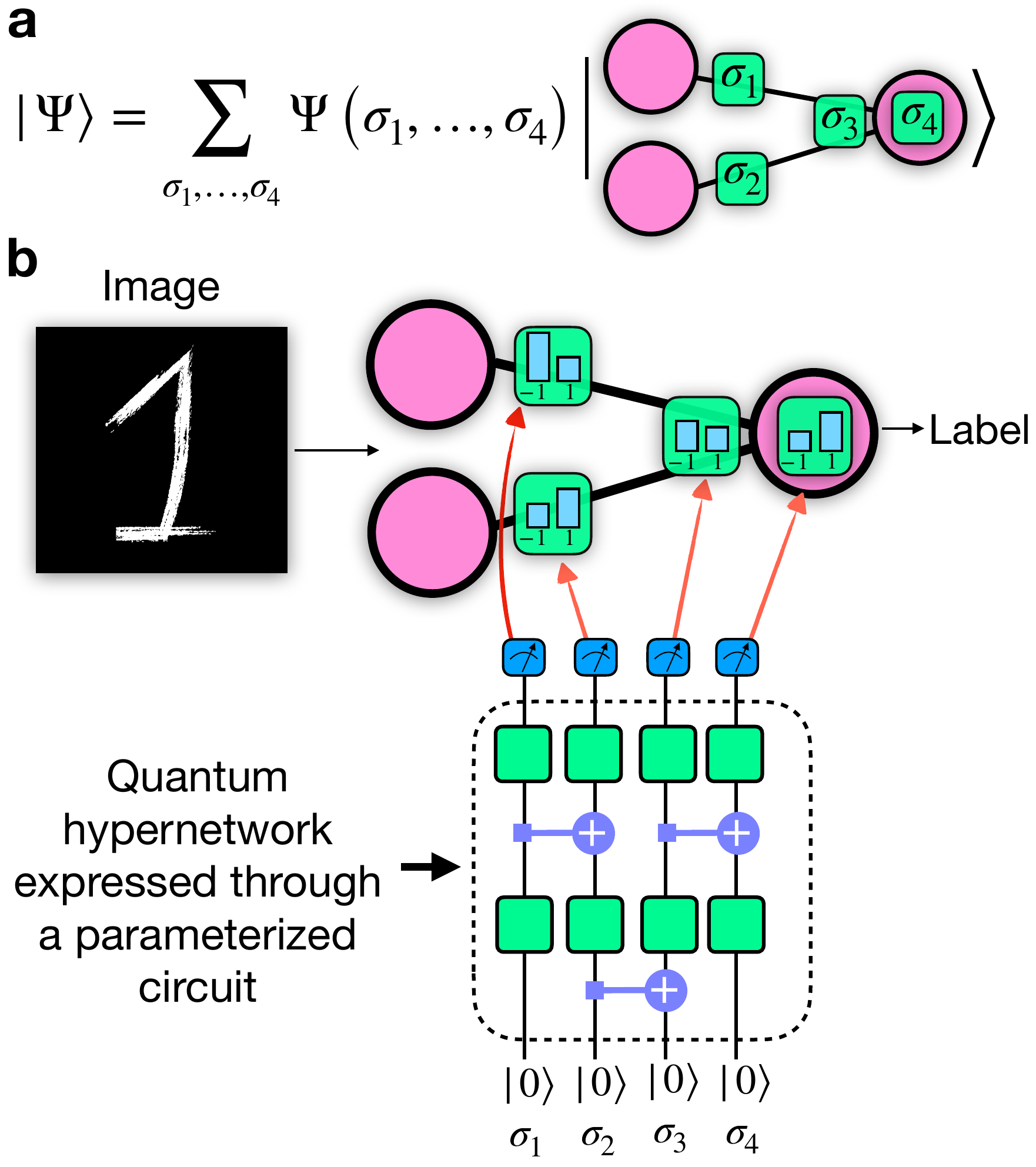} 
    \caption{{\bf Quantum HyperNetworks} (a) The different BiNN's configurations can be encoded in the computational basis $\boldsymbol{\sigma}$ of a quantum state $|\Psi\rangle$, which is defines a quantum hypernetwork. (b) The quantum hypernetwork can be constructed via a parameterized quantum circuit which upon measuring produces BiNN configurations. Different qubits $\sigma_i$ are interpreted as parameters, hyperparameters and architectural choices of the BiNN. }%
    \label{fig:encoding}%
\end{figure}

A quantum hypernetwork is a quantum state $|\Psi \rangle$ where each basis element $|\boldsymbol{\sigma} \rangle = |\sigma_1,\ldots,\sigma_N\rangle $ is associated with a specific configuration of an augmented model comprising the parameters of a BiNN, its hyperparameters, and any desired architectural choices to be encoded in the quantum hypernetwork. As quantum superposition is the feature of a quantum system whereby it exists in several separate states, i.e., all the different BiNNs encoded in Eq.~\ref{eq:qstate}, our approach can be understood as training BiNNs in quantum superposition. In Fig.~\ref{fig:encoding}(a) we represent a quantum hypernetwork encoding a small binary linear feed-forward network with a two-dimensional input and one-dimensional output. The BiNN is characterized by 2 weights (qubits $\sigma_{1}$ and $\sigma_{2}$), a bias (qubit $\sigma_{3}$), and an activation function. To encode architectural choices, e.g., the selection of activation function from two possibilities $f_1$ or $f_2$, we make the activation function qubit dependent (qubit $\sigma_4$ in  Fig.~\ref{fig:encoding}(a-b)), i.e., $f(x) \to f(x,\sigma)$, where, e.g.,  
\begin{equation*}
    f(\boldsymbol{x};\sigma) =
  \begin{cases} 
      f_1(\boldsymbol{x})& \text{if} \,\,\, \sigma=0 \\
      f_2(\boldsymbol{x}) &  \text{if} \,\,\, \sigma=1 . 
   \end{cases}
\end{equation*}
As explored below, other architectural choices and hyperparameters can be similarly encoded through the use of additional qubits. In this formulation, the number of qubits necessary to accommodate a problem with $N$ parameters and hyperparameters is $N$, i.e., linear in the size of the problem.  

A design principle for a quantum algorithm aiming at training classical neural networks may consist of the preparation of a quantum state $|\Psi \rangle$ (i.e. the hypernetwork) that assigns high amplitudes $\Psi(\sigma_1,\ldots,\sigma_N)$ to basis states $|\sigma_1,\ldots,\sigma_N\rangle$ encoding neural networks with a low cost function $C$ quantifying their performance 
\begin{equation}\label{eq:objective}
C\left(\boldsymbol{w} \right) = \frac{1}{N_{s}} \sum_{i=1}^{N_{s}}  \mathcal{L} \left( \text{NN}(\boldsymbol{x}_i;\{\boldsymbol{w} \}),\boldsymbol{y}_i \right).
\end{equation}
Here $N_s$ is the size of the training dataset composed of input $x_i$ and output $y_i$ variables, $\mathcal{L}$ is a loss function, and $\text{NN}(\boldsymbol{x}_i;\boldsymbol{w})$ represents an augmented neural network model. The augmented model parameters $\boldsymbol{w}=\{w_1,\ldots,w_{N}\}$, include the neural network weights, biases, hyperparameters, and architectural choices. The cost function corresponds to an $N$-bit real Boolean function  $C: \{0,1\}^{N}\to \mathbb{R}$, which the quantum algorithm aims to minimize. The weights and biases take the values $2\sigma_i-1 \in \{-1,1\}$.  

The simplest  approach to carry out this optimization using quantum resources is through a VQA. VQA employs a classical optimizer acting on a parameterized quantum circuit, with the purpose of finding solutions to a problem encoded in an objective function, which in our setting corresponds to $C\left(\boldsymbol{w} \right)$. A key element to a VQA is the encoding of the objective function, achieved by promoting Eq.~\eqref{eq:objective} to a quantum operator. A natural choice is to promote the parameters of the BiNN to a set of Pauli matrices  $\boldsymbol{w} \to \boldsymbol{\hat{\sigma}_{z}}=\left( \hat{\sigma}^{z}_1,\hat{\sigma}^{z}_2,\ldots, \hat{\sigma}^{z}_{N} \right)$, which, in turn, promotes the objective function $C\left(\boldsymbol{w} \right)$ to an operator $\hat{C}$. Here $\hat{\sigma}^{z}_i$ is the Pauli matrix $\begin{pmatrix}1&0\\0&-1\end{pmatrix} $ acting on qubit $i$, the diagonal cost operator 
\begin{equation}
 \hat{C}=\begin{pmatrix}
   C(\boldsymbol{w}_1) & 0 & \cdots & 0 \\
   0 & C(\boldsymbol{w}_2) & \cdots & 0 \\
   \vdots  & \vdots  & \ddots & \vdots  \\
   0 & 0 & \cdots & C(\boldsymbol{w}_{2^N})
 \end{pmatrix},    
\end{equation}
and $\boldsymbol{w}_1,\ldots,\boldsymbol{w}_{2^N}$ are all the $2^N$ possible BiNN. This encoding is flexible and other operator choices, including off-diagonal operators, are possible. 

We construct a quantum hypernetwork $|\Psi \rangle$ through a parameterized quantum circuit $U(\boldsymbol{\theta})$ with continuous parameters $\boldsymbol{\theta}$ such that $|\Psi \rangle \to |\Psi_{\boldsymbol{\theta}} \rangle =  U(\boldsymbol{\theta})|0\rangle^{\bigotimes n}$.  We aim at finding  solutions to the training of the BiNN solving for 
\begin{equation}\label{eq:optprob}
 \boldsymbol{\theta}^* = \operatornamewithlimits{\text{arg min}}_{ \boldsymbol{\theta}} \,\, E \left( \boldsymbol{\theta} \right),
\end{equation}
where $E \left( \boldsymbol{\theta} \right) = \langle \Psi_{\boldsymbol{\theta}} | \hat{C} |\Psi_{\boldsymbol{\theta}}\rangle$. From an ML perspective, this approach can be understood as a stochastic relaxation of the discrete optimization problem. That is, instead of directly searching for the optimal binary parameters, we introduce a joint distribution over the parameters and architectural choices encoded by the quantum state $ |\Psi_{\boldsymbol{\theta}} \rangle$. Measuring the quantum state (see Fig.~\ref{fig:encoding}(b)) produces trial binary parameters and architectural choices and gives access to estimates of the learning objective $E(\boldsymbol{\theta})$.

We express $U(\boldsymbol{\theta})$ as the product of $L$ unitary blocks of the form $U(\boldsymbol{\theta}) = U_{L}\left(\boldsymbol{\theta}_L\right )\cdots U_{1}\left(\boldsymbol{\theta}_1\right )$.  We restrict ourselves to one of the simplest and most widely available circuits in current quantum computing platforms, namely those implementable in quantum devices with a linear connectivity:  
\begin{align}
U_k\left(\boldsymbol{\theta}_k\right )&=\prod_{m=1+k\bmod{2},\text{ step } 2 }^{N-2+k \bmod{2} }\text{CX}(m,m+1) \\ \nonumber
&\prod_{j=1}^{N} \text{RY}(j,\theta_{y,j,k}) \text{RZ}(j,\theta_{z,j,k}).
\end{align}
Here $\text{CX}(m,j)$ denotes a control-X gate acting on the control $m$ and target $j$ qubits. The parameterized single-qubit unitaries $\text{RY}(j,\theta_{y, j, k})$  and $\text{RZ}(j,\theta_{z, j, k})$ at block $k$ are given by $e^{i \theta_{y, j, k} \hat{\sigma}^{y}_{j} }$ and $e^{i \theta_{z, j, k} \hat{\sigma}^{z}_{j} }$, respectively. The symbol $i$ is the imaginary unit. The parameters of the circuit are $\boldsymbol{\theta}= \{\theta_{\alpha, j, k} \}$, where $\alpha=y,z$, $j=1,\ldots,N$, and $k=1,\ldots,L$. We illustrate a quantum circuit with $L=2$ and $N=4$ in Fig.~\ref{fig:encoding}(b), where the green boxes synthesize the combined effect of  $\text{RY}(j,\theta_{y, j, k})$ and $\text{RZ}(j,\theta_{z, j, k})$. We note that a linear connectivity can be embedded, e.g., in heavy-hexagonal lattice. Out a heavy-hexagon lattice with 127 qubits, such as the one in the IBM Eagle processor~\cite{kimEvidenceUtilityQuantum2023}, it is possible to use 109 qubits arranged in a one dimensional fashion. In our experiments, we consider even $L=2\times N_{\text{layer}}$ and define a layer (see encircled blocks in Fig.~\ref{fig:encoding}(b)) as 2 unitary blocks, so that the circuit in Fig.~\ref{fig:encoding}(b) contains $N_{\text{layer}}=1$ layers. In addition, we also consider one of the simplest possible quantum states, namely an entanglement-free product state ansatz, where $U(\boldsymbol{\theta})_{\text{prod.}} = \prod_{j=1}^{N} \text{RY}(j,\theta_{y, j, 1}) \text{RZ}(j,\theta_{z, j, 1})$. The latter have been shown effective at solving quadratic unconstrained binary optimization problems~\cite{PhysRevA.104.062426,bowlesQuadraticUnconstrainedBinary2021}.

\subsection{Optimization}
We optimize the Eq.~\eqref{eq:optprob}  via a gradient-based method where $E\left( \boldsymbol{\theta} \right)$ and its gradient $\nabla_{ \boldsymbol{\theta}} E\left( \boldsymbol{\theta} \right)$ are evaluated through measuring the quantum hypernetwork $|\Psi_{\boldsymbol{\theta}}\rangle$ followed by a classical optimizer that iteratively updates its parameters. At the end of the optimization, we expect that  $|\Psi_{\boldsymbol{\theta}} \rangle$ assigns high amplitudes to BiNNs with low cost function, i.e., good architectural choices, parameters and hyperparameters. 

In an experimental setting, the estimation of the gradients  $\nabla_{ \boldsymbol{\theta}} E\left( \boldsymbol{\theta} \right)$ makes use of the parameter-shift rule~\cite{PhysRevA.98.032309,PhysRevA.99.032331}. 

It follows that the entries of the gradient are given by
\begin{equation}\label{eq:gradient}
  \frac{\partial  E \left( \boldsymbol{\theta} \right) }{\partial \theta_{\alpha, j, k} } = 
  \frac{1}{2}\left[ E(\boldsymbol{\theta}^{+}_{\alpha, j, k}) -  E(\boldsymbol{\theta}^{-}_{\alpha, j, k}) \right],
\end{equation}
where the elements of the shifted parameter vector $\boldsymbol{\theta}^{\pm}_{\alpha j k }$ are such that $\theta^{\pm}_{\beta, m, l} = \theta_{\beta, m, l} \pm \frac{\pi}{2}\delta_{\alpha,\beta} \delta_{m,j}\delta_{k,l}$. Thus, the calculation of the gradient corresponds to the evaluation of a shifted version of the objective function $E( \boldsymbol{\theta})$, which can be estimated by preparing and measuring the same quantum circuit used to compute the original objective with shifted circuit parameters. 

In a quantum experiment, functions of the form $E(\boldsymbol{\theta})$ are estimated via averages over the measurement outcomes of projective measurements, e.g.,
\begin{align}
        E \left( \boldsymbol{\theta} \right) &= \langle \Psi_{\boldsymbol{\theta}} | \hat{C}  | \Psi_{\boldsymbol{\theta}} \rangle \\ \nonumber
        &= \sum_{\sigma_1,\sigma_2,\ldots,\sigma_{N}} |\Psi_{\boldsymbol{\theta}}(\sigma_1,\sigma_2,\ldots,\sigma_{N})|^2 C(\sigma_1,\sigma_2,\ldots,\sigma_{N}) \\ \nonumber 
        &=\mathbb{E}_{\boldsymbol{\sigma} \sim |\Psi_{\boldsymbol{\theta}}|^2} \left[ C(\boldsymbol{\sigma})\right] \approx \frac{1}{N_{qc}} \sum_{i=1}^{N_{qc}} C(\boldsymbol{\sigma}_{i}),
\end{align}
where $N_{qc}$ configurations $\boldsymbol{\sigma}_{i}$ are distributed according to $ |\Psi_{\boldsymbol{\theta}}|^2$. The estimate of $ E(\boldsymbol{\theta}) \approx \frac{1}{N_{qc}} \sum_{i=1}^{N_{qc}} C(\boldsymbol{\sigma}_{i})$ is evaluated classically by computing $C(\boldsymbol{\sigma})$ on the BiNNs  $\boldsymbol{\sigma} \sim |\Psi_{\boldsymbol{\theta}}|^2$ sampled by the quantum computer.  
 
 In contrast, we use classical simulations based on tensor networks (TN)~\cite{osti_22403404} implemented through the PastaQ.jl package~\cite{pastaq}. PastaQ.jl relies on tensor-network representations of quantum states and processes. In particular, the quantum state is represented as a matrix product state for noise-free simulations. For noisy simulations, we use a matrix product operator representation of the density matrix of the system. The TN techniques allow for the exact evaluation of expectations and their gradients through automatic differentiation (AD) provided by the package Zygote.jl~\cite{innesDifferentiableProgrammingSystem2019}. The objective function $\hat{C}$ is constructed by fully enumerating all possible BiNNs, whose computational time scales exponentially in the number of variables of the problem. To optimize $E(\boldsymbol{\theta})$ we use the limited-memory Broyden–Fletcher–Goldfarb–Shanno (LBFGS) algorithm~\cite{10.5555/3112655.3112866}. Typical execution times of the classical simulation of the variational algorithms are provided in Appendix~\ref{ap:simt} as well as details about the execution time of the construction of $\hat{C}$. Implementations of our algorithms and datasets are available on the Github repository~\cite{GitHub2024}.

\subsection{Gaussian dataset with a choice of activation} 
\begin{figure}%
    \includegraphics[width =\linewidth]{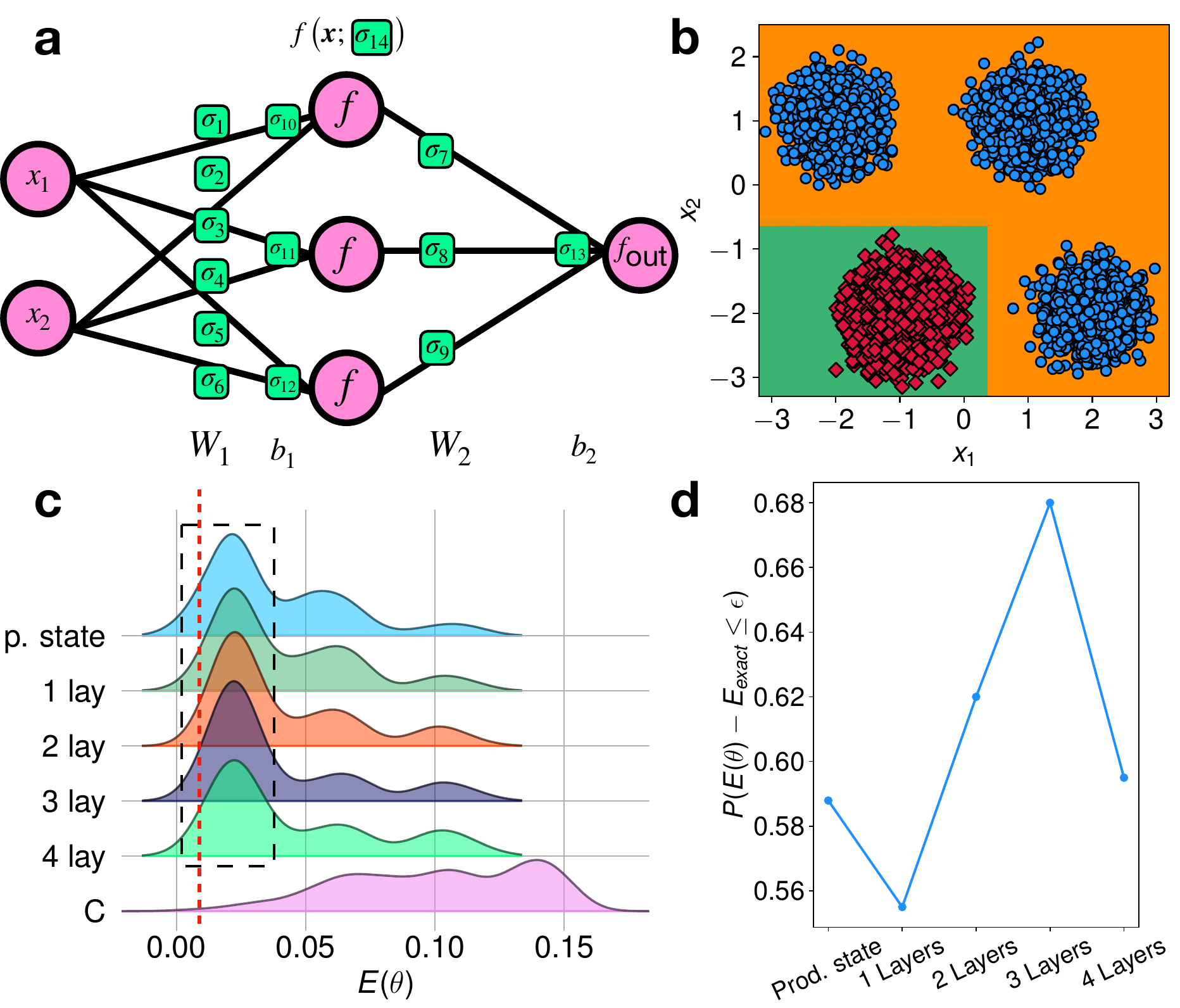}
    \caption{ \textbf{BiNNs applied to a Gaussian dataset.} (a) A BiNN with two dimensional input, a three dimensional hidden layer, and one output. (b) The decision boundary drawn by our BiNN after training on a two-dimensional mixture of 4 Gaussian distributions with two labels. (c) A kernel density estimation (KDE) of the probability that a quantum circuit (product state and with different number of layers)  achieves an average cost $E(\theta)$. The bottom row of this panel corresponds to the density of configurations with a cost $C(\bm{w})$ after filtering for low-cost configurations. (d) The probability of finding the lowest objective $C(\bm{w^*})$ within $\epsilon=0.03$ for the different quantum circuits.}
    \label{fig:gaussian}%
\end{figure}

We first consider the training of a small BiNN binary classifier with a two-dimensional input, a three-dimensional hidden layer, and a single output depicted in Fig.~\ref{fig:gaussian}(a). We would like to simultaneously train the parameters, as well as an architectural choice, here the selection of activation function $f$, which in our example can be a sigmoid or a rectified linear unit (ReLU):
\begin{equation*}
    f(\boldsymbol{x};\sigma) =
  \begin{cases} 
      S(\boldsymbol{x})& \text{if} \,\,\, \sigma=0 \\
      \text{ReLU}(\boldsymbol{x}) &  \text{if} \,\,\, \sigma=1. 
   \end{cases}
\end{equation*}
Here  $S(\boldsymbol{x})=1/(1+e^{-\boldsymbol{x}})$ and  $ \text{ReLU}(\boldsymbol{x})=\text{max}(\boldsymbol{x},0)$ are applied element-wise on the components of the arrays $\boldsymbol{x}$. The activation function in the output layer is  $f_{\text{out}}(\boldsymbol{x})=S(\boldsymbol{x})$.  

We train the BiNN on a toy dataset drawn from a two-dimensional mixture of 4 Gaussian distributions shown in Fig.~\ref{fig:gaussian}(b). The samples are drawn from the red (squares) Gaussian with probability $1/2$ and from each of the blue (circles) Gaussians with probability $1/6$. Each data point is labeled according to whether it was drawn from the red or blue Gaussians, and we aim to train the BiNNs to classify any point in the plane accordingly.   

The BiNN is characterized by 13 binary parameters and a binary variable $\sigma_{14}$ codifying the architectural choice of activation function, i.e. $N=14$ variables. For small BiNNs, a training dataset, and an objective function $C\left(\boldsymbol{w} \right)$, it is possible to compute the optimal BiNN configuration by enumerating all the $2^N$ BiNNs and choosing the one with the smallest $C\left(\boldsymbol{w} \right)$. In our example, the best configuration yields the decision boundary shown in Fig.~\ref{fig:gaussian}(b), where the optimal BiNN classifies points in the green region as coming from the red Gaussian and points in the orange region  as coming from the blue Gaussians. The optimal choice of activation function is the ReLU. %
Now we explore solving the problem via quantum optimization. We consider the circuit ansatz shown in Fig.~\ref{fig:encoding} with varying number of layers $N_{\text{layer}}=1,2,3,4$, as well as a product state ansatz. We randomly and independently initialize all the parameters of the circuits from a uniform distribution $\theta_{\alpha,i,k}\sim U(0,2\pi)$. We first note that all the circuit ansatze have sufficient expressive power to represent the optimal solution, which is simply the product state. In our numerical experiments, we find that all of our ansatze can find the optimal solution, including the product state ansatz, with varying degree of success. The success rate of the optimization depends on the interplay between the initialization of the circuit parameters and the depth of the circuit. To understand the typical behaviour of the optimization procedure and to shed light onto the role of the circuit depth, we perform the circuit optimization for a number $N_{\text{optim}}= 200$ of independent initializations.

In Fig.~\ref{fig:gaussian}(c) we summarize the results of these optimizations through a kernel density estimation (KDE) of the probability density function that the optimization finds an average objective $E(\boldsymbol{\theta})$ for different circuit depths. 
In addition, through full enumeration, we compute a KDE of the 200 top performing BiNN with the lowest $C(\boldsymbol{w})$ (the bottom row of Fig.~\ref{fig:gaussian}(c)).
This can be interpreted as the density of configurations at a particular ``$E$ level''  that a BiNN can take, and is analogous to the density of states in condensed matter physics. Since we only take the 200 lowest objective function BiNNs, this means that the probability assigned by the KDE to each value $C(\boldsymbol{w})$ is significantly overestimated. 

We observe that most solutions found by all circuits considered here are concentrated near the optimal configuration of weights and hyperparameters $\boldsymbol{w}^{*}$, for which $C(\boldsymbol{w}^{*})\approx 0.008$ (see the encircled densities in Fig.~\ref{fig:gaussian}(c)). This is in spite of the fact that the density of solutions with low $C$ is significantly small, which indicates that the quantum optimization is effective. However, the frequency with which solutions with low objective function are found varies as a function of the circuit depth. To probe this behaviour, we estimate the probability that a certain circuit depth finds solutions with precision $E(\boldsymbol{\theta}) -C(\boldsymbol{w}^{*}) < \epsilon$ by counting the solutions found by the VQA meeting the precision condition. This is shown in Fig.~\ref{fig:gaussian}(d). As noted earlier, even a product state circuit finds accurate solutions. A circuit with 1 layer (see Fig.~\ref{fig:encoding}) doubles the number of variational parameters and decreases the probability to find accurate solutions which indicates that the optimization of the variational parameters $\boldsymbol{\theta}$ is more prone to getting stuck in local minima than a product state circuit. Upon increasing the depth, we note that for 2 and 3 layers, the probability to find accurate solutions reaches a maximum but eventually decreases for 4 layers. Circuits with layers composed of entangling gates and multiple (optimally for $N_{\text{layer}}=3,4$) layers of parameterized single-qubit gates provide an advantage as these enhance the success of finding good solutions with respect to a product state.    
\begin{figure}%
    \includegraphics[width =\linewidth]{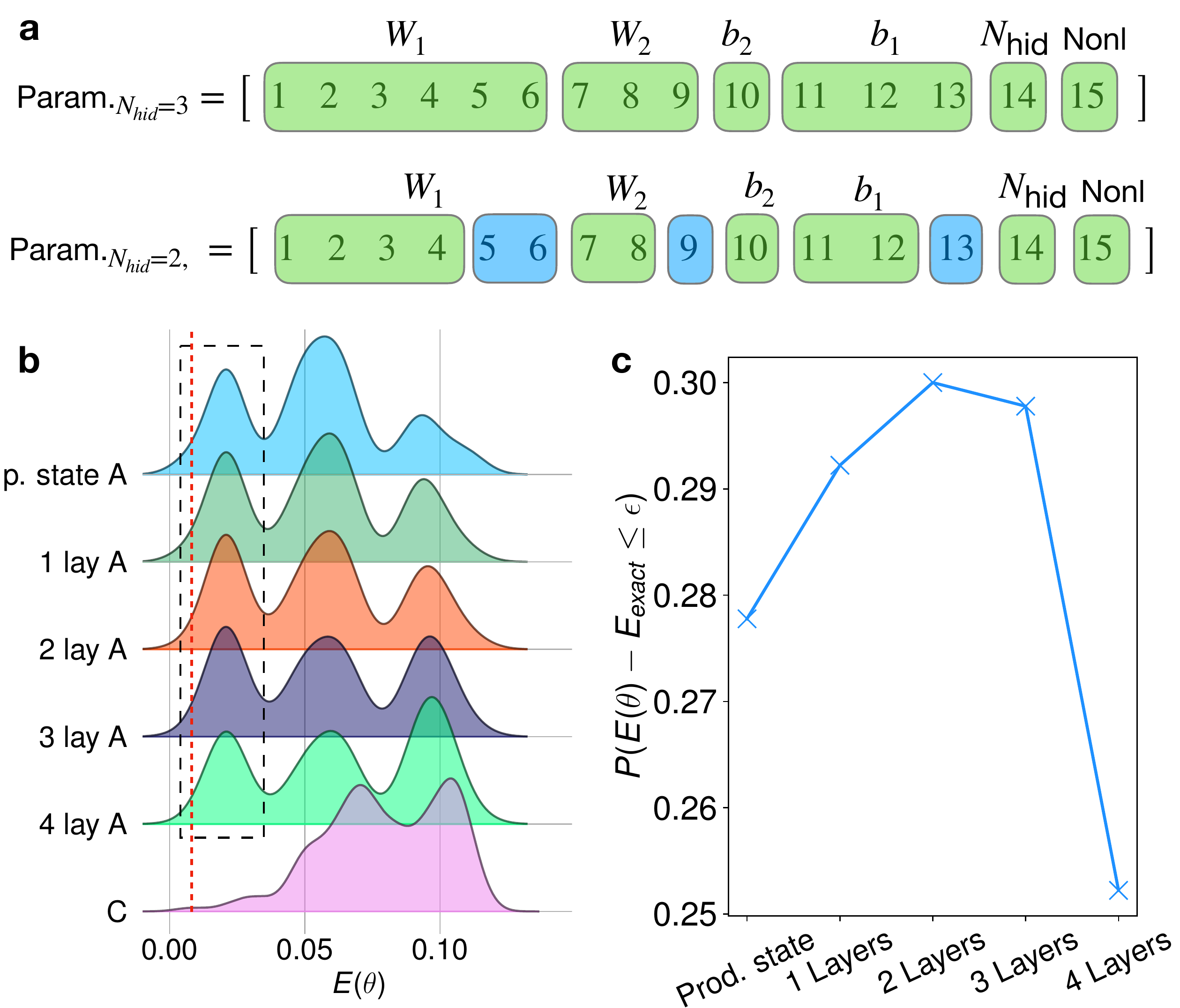} 
    \caption{\textbf{BiNNs with two layers with width and nonlinearity hyperparameters on the Gaussian toy dataset.} (a) An illustration of the weights, biases, hyperparameters within a flattened list for two different numbers of hidden neurons $N_{hid} = 2, 3$. Blue color means that the corresponding parameter is not used. (b) Similarly to Fig.~\ref{fig:gaussian}(c), we show the KDE for different quantum circuits' objectives in the top five rows, as well as the density of configurations in the bottom row. (c) In a similar fashion to Fig.~\ref{fig:gaussian}(d), we show the probability of success within a threshold $\epsilon=0.03$ for each quantum circuit.}%
    \label{fig:gaussian_Nhid_nonl}
    
\end{figure}

\subsection{Gaussian dataset with a choice of activation and dimension of hidden layer }
Next we consider the simultaneous optimization of the BiNN’s parameters, a hyperparameter (hidden layer dimension $N_{hid}$) and the architectural choice non-linearity. We encode the choice of $N_{hid} \in \{2,3\}$ through an additional qubit $\sigma_{N_{hid}}$. A wider set of choices of $N_{hid}$ is possible through the use of more qubits. We encode these choices through a single function $\text{NN}(\boldsymbol{x}_i;\boldsymbol{w})$ that evaluates the BiNN's output as a function of weights and biases, and the choices of non-linearity and $N_{hid}$. The choice of qubit assignment of the BiNN's parameters and nonlinearity are presented in Fig.~\ref{fig:gaussian_Nhid_nonl}(a), where the qubits encircled in blue are left unused in the evaluation of the BiNN output for $N_{hid}=2$ but are used for $N_{hid}=3$. 

The results of the optimization procedure are displayed in Fig.~\ref{fig:gaussian_Nhid_nonl}(b-c), which display a behaviour similar to the experiments in  Fig.~\ref{fig:gaussian}(c-d).  
While the optimization is successful, the probability of finding low-energy solutions is reduced with respect to the original optimization task in Fig.~\ref{fig:gaussian}. As noted in Fig.~\ref{fig:gaussian},  while even a product state circuit finds accurate solutions with high probability, there exists an optimal  circuit depth that significantly enhances the probability of a finding the optimal solution, eventually decreasing upon further increasing depth. This effect is due to the optimization becoming more prone to finding local minima and not to the ansatz’ expressive power, as deeper circuits are more expressive than shallower ones.   

\begin{figure}%
   
    \includegraphics[width =\linewidth]{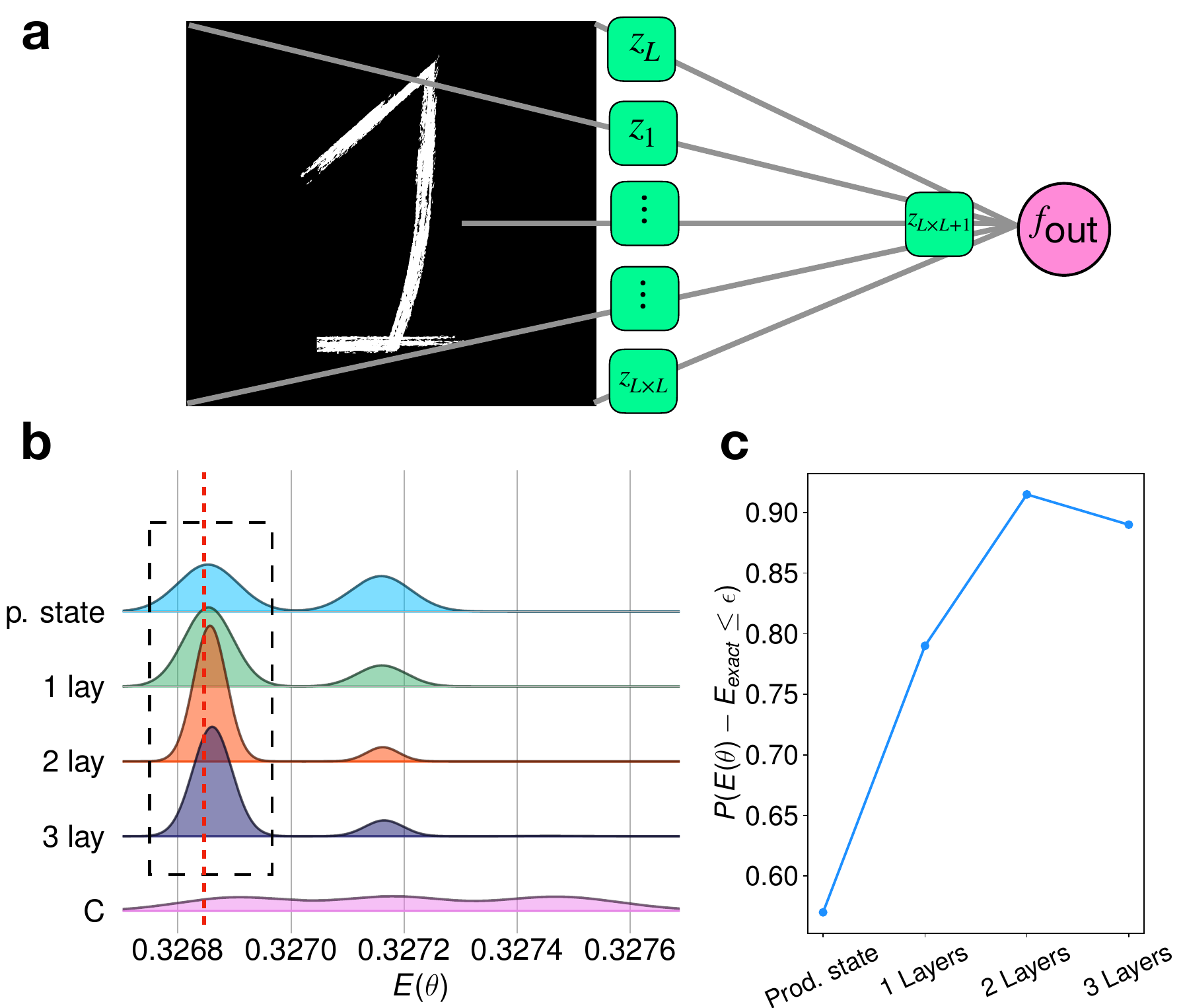} 
    \caption{\textbf{BiNNs for the reduced MNIST dataset.} (a) The BiNN used to classify the reduced MNIST dataset. (b) A KDE of $E(\boldsymbol{\theta})$ resulting from repeating the optimization procedure 200 times. The optimizations are performed for a product state, as well as for circuits with 1,2, and 3 layers. We also show the density of configurations with a cost $C(\bm{w})$. (c) A plot of the probability of success within a threshold $\epsilon=5\times10^{-5}$ for the different quantum circuit architectures.}%
    \label{fig:MNIST}
\end{figure}
\subsection{Scaled-down MNIST}
To investigate the performance of quantum optimization for a representative dataset, we consider training a simple logistic regression model with binary parameters and a cross-entropy loss, as shown in Fig.~\ref{fig:MNIST}(a). The training data corresponds to a subset of MNIST images of zeros and ones scaled down to $L\times L = 4\times 4 $ pixels. We explore solving the training problem via quantum optimization with the circuit depths $N_{\text{layer}}=1,2,3$ as well as with a product state circuit. As in our previous example, we run the optimization for $N_{\text{optim}}= 200$ independent initializations. The results are shown in  Fig.~\ref{fig:MNIST}(b). We find that the best results for MNIST are obtained by mapping the weights and biases to the values $\sigma_i \in \{0,1\}$. 

While optimization via a product state ansatz attains optimal solutions with nearly $60\%$ success rate, the rate is enhanced for circuits with an optimal number of entangling layers, which display a $90\%$ chance of success for $N_{\text{layer}}=2,3$. This backs up our previous observation that entanglement plays an important role in the optimization procedure. As seen in Fig.~\ref{fig:MNIST}(b-c) our circuits enhance the probability of finding the optimal solution going beyond simply increasing of the expressive power of the circuit ansatz. 

\begin{figure}%
   
    \includegraphics[width =\linewidth]{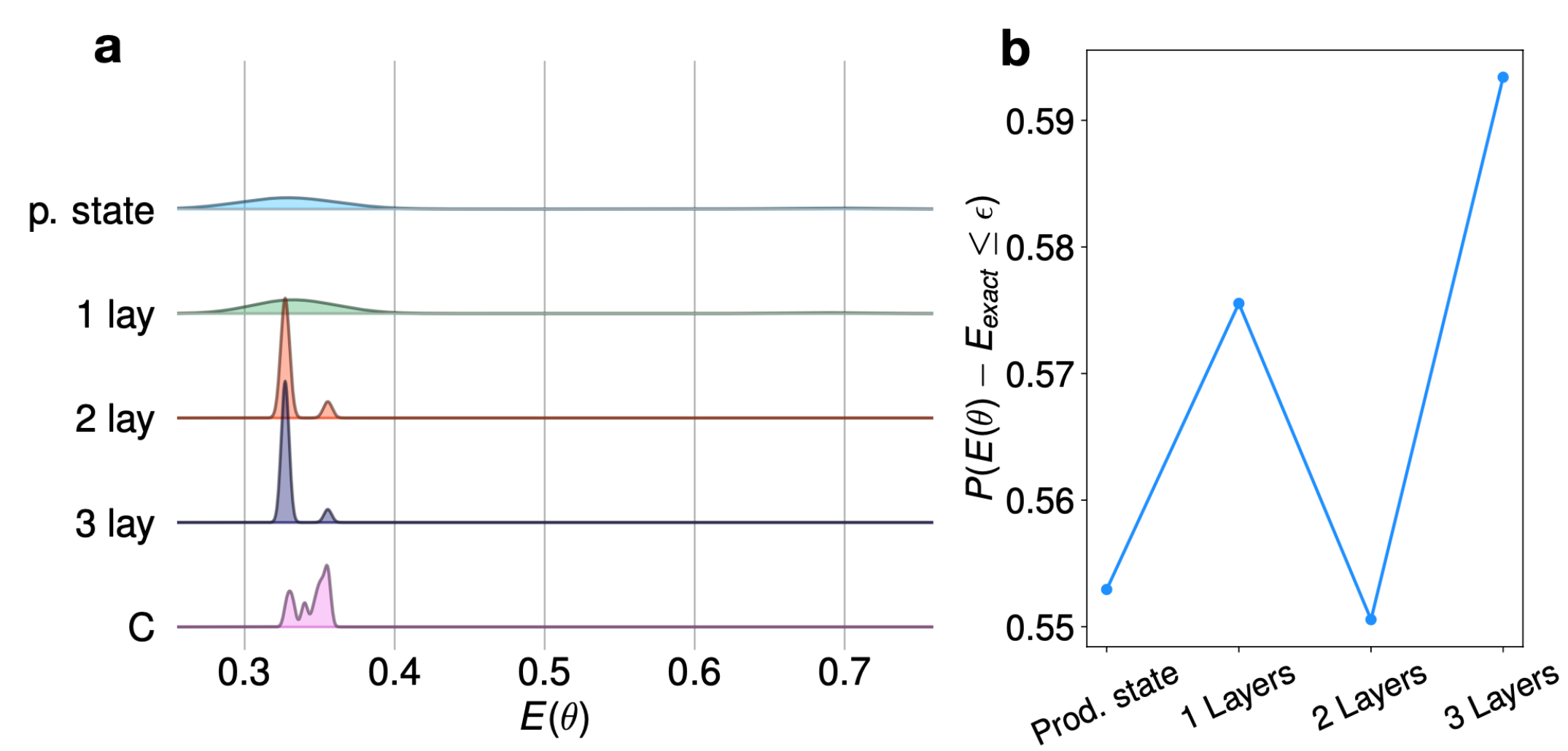} 
    \caption{\textbf{Encoding and parameter search for the reduced MNIST dataset.} (a) A KDE of $E(\boldsymbol{\theta})$ resulting from repeating the optimization procedure 200 times for the MNIST dataset with encoding search. (b) A plot of the probability of success within a threshold $\epsilon=1\times10^{-4}$ for the different quantum circuit architectures.}%
    \label{fig:MNISTENCODING}
\end{figure}

Finally, while for the Gaussian dataset we have mapped the weights and biases to $2\sigma_i-1 \in \{-1,1\}$, in the MNIST example we have mapped them to $\sigma_i \in \{0,1\}$. However, it is possible for our algorithm to perform a search over multiple encodings in superposition. In Fig.~\ref{fig:MNISTENCODING} we demonstrate numerical simulations for our algorithm searching over bias, weights, encoding choices, namely mapping weights and biases to binary values in $\{0,1\}$, $\{-1,1\}$, $\{-2,1\}$, and $\{-3,1\}$. The search over such an additional space of encodings is carried out by adding two additional qubits accounting for the 4 different possible encodings. Both the KDE and probability of success behave similarly to our other experiments where additional depth is seen to contribute to the success of the optimization procedure.  

\subsection{Fourier analysis of $C$}
In spite of the similarities between the MNIST and Gaussian mixture examples in terms of problem size $N$ and task, we note that the probability of finding solutions with low cost function is higher for the MNIST task. To shed light onto the origin of these differences in optimization performance, we examine the structure of the objective functions $C$ through a Fourier analysis.

As pointed out by~\citet{tortaQuantumApproximateOptimization2021}, due to the non-linearities of the BiNNs and the loss function $\mathcal{L}$, the objective function $C$ and its quantum extension $\hat{C}$  may contain highly non-local, all-to-all multi-variable interactions. Beyond understanding the differences in optimization performances across different tasks, the locality of $C$ plays an important role in the optimization of the circuit as a highly non-local $C$ may lead to exponentially vanishing gradients in Eq.~\ref{eq:gradient}, which can severely impede the optimization of the circuit~\cite{cerezoCostFunctionDependent2021}.

The Fourier transform of the real boolean function $C$ and its quantum extension $\hat{C}$ provides a natural strategy to investigate the locality of the objective function $C$. First, $\hat{C}$ can be represented by an Ising Hamiltonian given by sums of tensor products of Pauli $\sigma^z_i$ operators weighted by $C$'s Fourier expansion coefficients~\cite{10.1145/3478519}. Thus, for an  $N$-bit real function  $C: \{0,1\}^{N}\to \mathbb{R}$, we can decompose $\hat{C} |\mathbf{\sigma}\rangle= C(\mathbf{\sigma})|\mathbf{\sigma}\rangle$ as 
\begin{equation}
   \hat{C} = \sum_{\hat{\sigma}_1, \ldots \hat{\sigma}_N } f(\hat{\sigma}_1, \ldots \hat{\sigma}_N) \bigotimes_{i=1}^{N} \hat{\sigma}_i,
\end{equation}
where $\hat{\sigma}_i=\{ \boldsymbol{1}, \hat{\sigma}_i^{z} \}$. Here the Fourier coefficients are given by $f(\hat{\sigma}_1, \ldots \hat{\sigma}_N) = \frac{1}{2^N}\text{Tr}\left[ \hat{C} \otimes_{i=1}^{N} \hat{\sigma}_i \right] \in \mathbb{R}$. This follows from the fact that the tensor products of Pauli operators and the identity form an orthogonal basis for the vector space of $2^N\times2^N$ complex matrices, in particular the subspace of diagonal operators such as $\hat{C}$, for which only the $2\times 2 $ identity matrix $\boldsymbol{1}$ and $\sigma^z_i$ are required in the expansion.      

We evaluate the $N$-bit function $f(\hat{\sigma}_1, \ldots \hat{\sigma}_N)$ and define the amplitude
\begin{equation}
W(S)= \sum_{\hat{\sigma}_1, \ldots \hat{\sigma}_N } |f(\hat{\sigma}_1, \ldots \hat{\sigma}_N)|^2   \delta_{S, S(\hat{\sigma}_1, \ldots \hat{\sigma}_N)}. 
\end{equation}
 as the total sum of the Fourier coefficients squared associated with diagonal Pauli strings with weight $S$. Here the weight $S(\hat{\sigma}_1, \ldots \hat{\sigma}_N )\in \{ 0..N \}$ of an $N$-length Pauli string $\bigotimes_{i=1}^{N} \hat{\sigma}_i$ corresponds to the number of non-identity Pauli matrices in it. 

The structure of the function $W(S)$ reflects the locality of the effective Ising Hamiltonian. For instance, when $W(S)\neq0$ only for $S=0,1$  means that the effective Ising Hamiltonian corresponds to a set of local fields acting independently on the variables $\sigma_i$. In contrast, if $W(S)\neq 0$ for $S=0,1,2$ means that the Ising Hamiltonian contains only pairwise interactions and local fields, etc.  
Speaking informally, $W(S)$ defines how well we can approximate $\hat{C}$ with a polynomial of degree $S$.

\begin{figure}%
    \includegraphics[width =\linewidth]{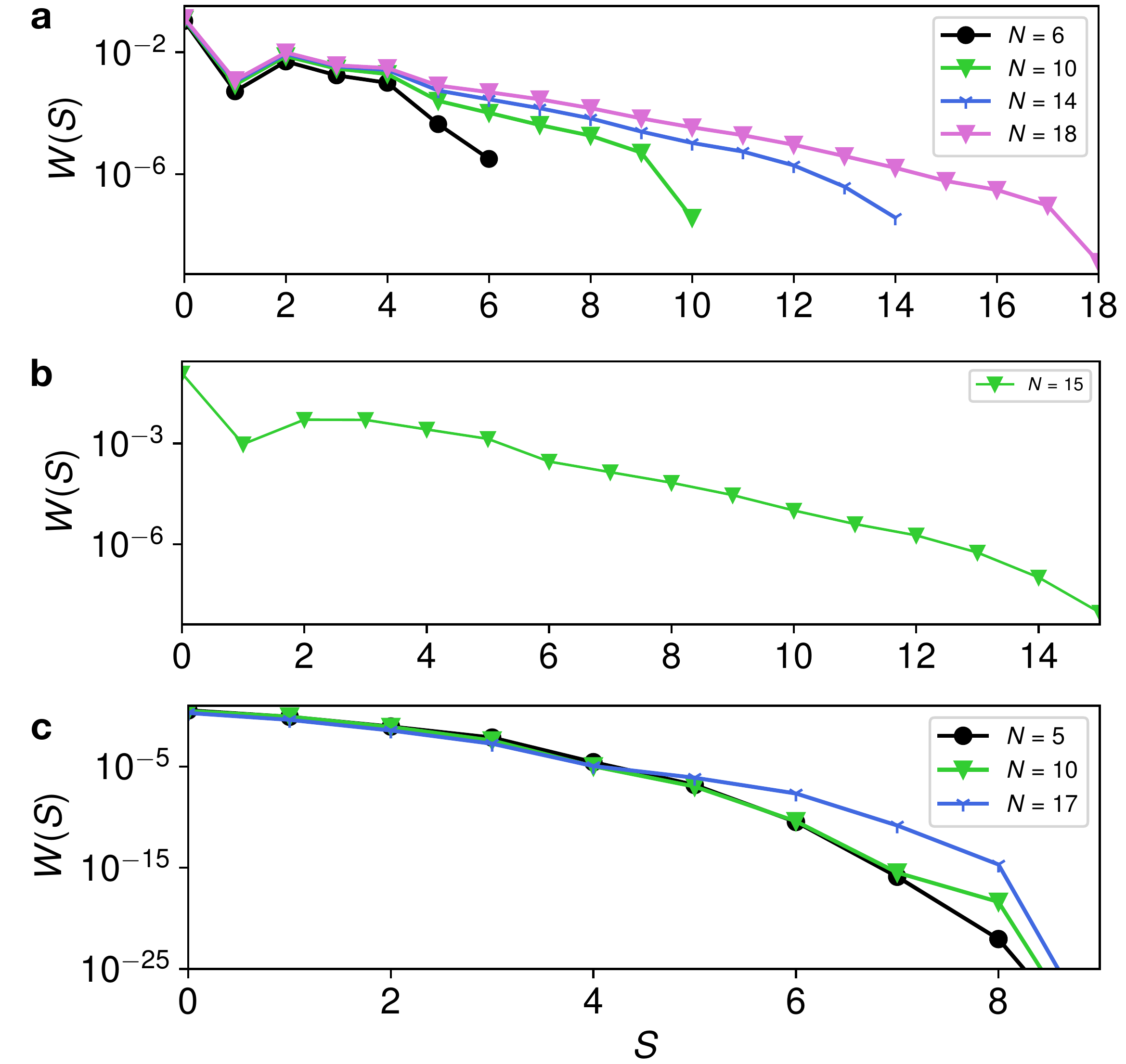} 
    \caption{\textbf{Fourier analysis of $\hat{C}$}. The amplitudes $W(S)$ associated with Pauli string weights $S$ for (a) the classification of Gaussians with activation function search, (b) the classification of guassians with the hidden neurons and activation function search, (c) the logistic regression of reduced MNIST. }%
    \label{fig:fourier}%
\end{figure}

In Fig.~\ref{fig:fourier}(a-c) we show $W(S)$ for the tasks of classification of Gaussians with function activation search (a), Gaussians with activation function and hidden dimension search (b), and logistic regression of MNIST with binary weights (c). For all systems, the highest $W(S)$ happens at $S=0$, which corresponds to a simple constant shift in the effective Hamiltonian. As the weight $S$ increases, the amplitude $W(S)$ is seen to decrease exponentially fast even for moderate $S$. This means that all the objectives $C$ are essentially local, which bodes well for circuit optimization as the locality of $C$ will not induce barren plateaus~\cite{cerezoCostFunctionDependent2021}. For the Gaussian mixture tasks, we see that the most important contributions to $W$ occur at $S=2,3$ with the highest values occurring at $S=2$, which means that the effective Hamiltonian is nearly an Ising Hamiltonian with pairwise interactions. Instead, for MNIST the most dominant non-trivial contribution comes from $S=1$, which means that the effective Hamiltonian is a set of local fields acting on the binary weights of the model. This in part explains why the quantum optimization of the MNIST task is superior since the solution of a fully independent set of binary variables coupled to local fields can be found by independently optimizing the energy of each binary variable. This means that a product state is perfectly suited to find it with high probability, as we have found. Additionally, these observations support the idea that short-depth circuits with one- and two-qubit gates can tackle the optimization of the BiNNs without resorting to full implementations of unitaries of the form $e^{i\hat{C}}$, which have been typically prescribed in earlier proposals for training neural networks using quantum computers~\cite{tortaQuantumApproximateOptimization2021,Verdon2018}.   

\subsection{Overfitting and model selection}

In contrast with the standard ML workflow
where the hyperparameter and architectural choices are optimized based on the model's performance on a validation set, 
we have defined an augmented model encapsulating the parameters, hyperparameters and architecture of a neural network, which we jointly optimize on a training dataset. We now briefly explore the generalization and overfitting behaviour of the augmented model and investigate how the resulting trends can guide the selection of an optimal augmented model. 

As an example, we revisit the training of a BiNN with architectural choice of non-linearity to the mixture of Gaussians dataset as a testbed to explore overfitting and generalization. To amplify overfitting, we simultaneously bring the Gaussians spatially closer to each other with respect to the example in Fig.~\ref{fig:gaussian} and decrease the size of the training dataset. As a function of the training iterations, we investigate the behaviour of $\overline{E(\boldsymbol{
\theta})}$, which is an average of $E(\boldsymbol{
\theta})$ over 100 independent realizations of training datasets of sizes $N_s =2,4,6,8$. We also consider training datasets with $N_s=10^{3}$, significantly larger than the dimensionality of the input ($d=2$). We fix the total number of circuit training iterations to $50$ and the choose a large validation set of size $1000$.   

Overall, we find that the augmented model adheres to the anticipated behaviour of an ML model. In all of our examples, the average training curves on the training sets are monotonically decreasing. For small training sets, e.g. $N_s=2$, the validation set curve initially decreases and later on increases, which suggests that the simple ``early stopping'' strategy may be employed to choose optimal models located at the minimum of the validation curve as a way to avoid overfitting~\cite{yaoEarlyStoppingGradient2007a}. For $N_s>2$, the validation curves exhibit a monotonic decreasing behaviour as a function of training iterations, which is the typical dynamics for large training sets $N_s$ where the dynamics is less prone to overfitting.  As expected, the generalization gap, i.e. the difference between the validation and training set curves near the end of the training, decreases quickly as a function of the $N_s$ and is seen to grow small for large datasets $N_s=10^{3}$, as expected.

\begin{figure}%
     \includegraphics[width =\linewidth]{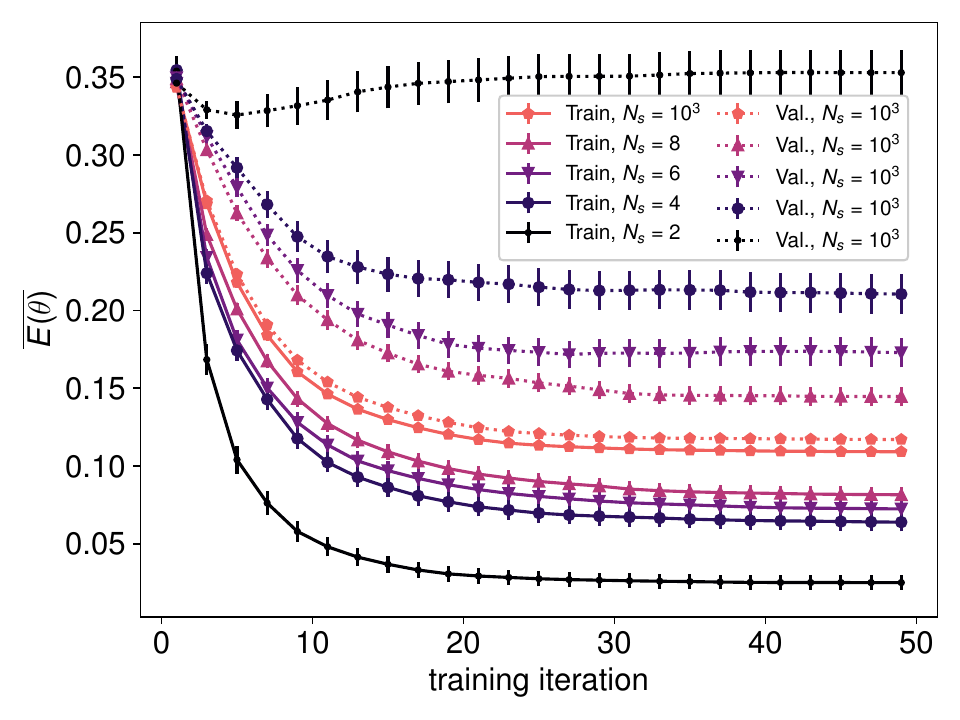} 
    \caption{ \textbf{Augmented model selection and overfitting.} Average behaviour of $E(\boldsymbol{\theta})$ over multiple realizations of the training datasets as a function of training iterations and size of the dataset $N_s$. We show the average $\overline{E(\boldsymbol{\theta})}$ computed on the training (solid lines) and validation sets (dotted lines). }
   
    \label{fig:overfitting}
\end{figure}

\subsection{Impact of local depolarization noise and gradient-free optimization.} 

To understand the robustness of our experiments to noise, we have performed simulations of our algorithm in the presence of local depolarization for the MNIST experiments. We assume a local depolarization following the application of each gate, including both single-qubit and two-qubit gates. 

We explore the impact of increasing levels of noise on performance across different depths for the MNIST problem. Our findings suggest that noise significantly influences the behavior of $E(\theta)$, adversely affecting the method's average success rate. Motivated by this observation, we have extended our analysis to examine the distribution of solutions sampled from the final state. In contrast with noise-free simulations where the distributions over configurations are strongly peaked near the basis element $\sigma$ that minimizes $E(\theta)$, we find that noisy simulations broaden the quantum state's probability distribution over computational basis states significantly, naturally leading to an increase of the optimal $E(\theta)$.  

Fortunately, despite the optimal $E(\theta)$ being notably higher than its noise-free counterpart, our experiments consistently show that the optimal BiNN is reliably found in the final quantum state with high probability. This indicates that, despite noise, we can efficiently explore a range of high quality solutions sampled by a noisy device. 

Similarly, we conducted simulations employing gradient-free optimization techniques, as well as both noisy and gradient-free optimization. Our overall finding indicates that the behavior of $E(\theta)$ is impacted by optimization without gradients. We also explore the distribution of solutions sampled from the final states. Despite $E(\theta)$ being notably higher than its gradient-full counterpart, our experiments consistently demonstrate that the optimal BiNN appears in the final quantum state with high probability.

We note that, among the quantum circuits explored in this work,  the least impacted by the presence of noise and gradient-free optimization is the product state. This implies that the availability of gradients remains crucial for the success of the method for circuits beyond simple product states. Similarly, this suggests that the levels of hardware noise should be sufficiently low so that the advantageous effects brought by circuit depth seen in our experiments are not washed out by noise.     

Numerical results and technical details about the noisy and gradient-free optimization simulations are presented in Appendices \ref{ap:gradfree} and \ref{ap:locdep}.

\section{Discussion}
We have introduced quantum hypernetworks, variational quantum circuits that search for optimal BiNNs over an augmented space comprising its parameters, hyperparameters, and any desired architectural choices, in a single optimization loop. Using classical simulations, we have shown that quantum hypernetworks effectively find optimal parameters, hyperparmeters and architectural choices with high probability on toy classification problems including a two-dimensional Gaussian dataset and a scaled-down version of the MNIST handwritten digits. We find that the probability of finding performant BiNNs is tied to the circuit depth, and consequently, to the amount of entanglement supported by the circuit. This indicates that entanglement and quantum effects play a role and decrease the probability that the optimization finds poor local minima.

Even though expressing quantum hypernetworks in terms of circuits with simple linear connectivity has proven successful in our setting, other ansatzes constructed considering knowledge of the problem, e.g., circuit designs adaptively grown guided by the objective function and gate availability~\cite{grimsleyAdaptiveVariationalAlgorithm2019}, may simultaneously shorten the circuit depth and significantly improve the effectiveness and scalability of our approach. To explore training large models beyond what's feasible with limited-size quantum processors, it is natural to consider a layer-by-layer optimization of the BiNN, which would operate analogously to the density matrix renormalization group algorithm~\cite{PhysRevB.48.10345}. Additionally, distributed quantum computing~\cite{thamQuantumCircuitOptimization2022}, as well as a multi-basis encoding of the problem~\cite{pattiVariationalQuantumOptimization2022,sciorilliLargescaleQuantumOptimization2024}, may extend the scalability of our approach to larger ML models. For instance, using a multi-basis encoding, the optimization of a binary neural net with $O(10^6)$ parameters, which reaches the scale of current BiNNs such as binary Resnet-18~\cite{huangComprehensiveBenchmarkingBinary2023}, would require $O(100)$ qubits using with a cubic root scaling~\cite{sciorilliLargescaleQuantumOptimization2024}. Another powerful strategy that may help scale the size of the problems amenable to quantum optimization is through the use of mid-circuit measurement and reset strategies~\cite{decrossQubitreuseCompilationMidcircuit2022} including the quantum matrix product state technique~\cite{PhysRevResearch.1.023025,PhysRevB.106.165126,PhysRevLett.128.150504}. The application of fault-tolerant quantum algorithms may also prove useful to the success of the unified training strategy presented here and may lead to provable speedups~\cite{liaoQuantumSpeedupGlobal2021}.

Our approach naturally connects with Bayesian inference as the quantum hypernetwork $|\Psi_{\boldsymbol{\theta}}\rangle$ defines a probability distribution over the weights of the BiNNs, a defining property of a Bayesian neural network. A full Bayesian approach prescribes the evaluation of the posterior distribution over the parameters, which is fundamentally intractable in our setting~\cite{PhysRevApplied.16.044057,nikoloskaQuantumAidedMetaLearningBayesian2022}. It may be possible, however, to estimate the evidence lower bound~\cite{jordan1999learning} by performing a decomposition of the circuit distribution, taking inspiration from~\citet{titsias2019unbiased,pmlr-v89-molchanov19a}. Additionally, although we have arrived at our unified strategy through a variational quantum optimization lens, our approach suggests that it is possible to introduce classical or quantum-inspired hypernetworks based on, e.g., recurrent neural networks or product states~\cite{VNA2021,bowlesQuadraticUnconstrainedBinary2021}, where a variational Bayesian approach to BiNN optimization is possible.

Quantum computers are currently reaching the ability to vastly outperform supercomputers' energy efficiency by many orders of magnitude over classical computers~\cite{villalongaEstablishingQuantumSupremacy2020}, so it stands to reason that the efficiency and energetic consumption of complex tasks in the ML workflow such as neural network training and hyperparameter search may be significantly reduced through the use of quantum computational resources. A combination of the energy efficiency of BiNN's classical operation with the energetic advantages of quantum devices for their training along with the unified single-loop optimization introduced here may offer a compelling approach to train large ML models with a reduced carbon footprint in the future.

\section*{Acknowledgments}
We acknowledge Maciej Koch-Janusz, Roeland Wiersema, Roger Melko, and Behnam Javanparast for discussions. JC acknowledges support from the Natural Sciences and Engineering Research Council (NSERC), the Shared Hierarchical Academic Research Computing Network (SHARCNET), Compute Canada, and the Canadian Institute for Advanced Research (CIFAR) AI chair program. Resources used in preparing this research were provided, in part, by the Province of Ontario, the Government of Canada through CIFAR, and companies sponsoring the Vector Institute \url{www.vectorinstitute.ai/#partners}. Research at Perimeter Institute is supported in part by the Government of Canada through the Department of Innovation, Science and Economic Development Canada and by the Province of Ontario through the Ministry of Economic Development, Job Creation and Trade. 



\bibliography{Biblio}

\appendix 

\section{Appendix: Gradient-free optimization}\label{ap:gradfree}

\begin{figure}%
   
    \includegraphics[width =\linewidth]{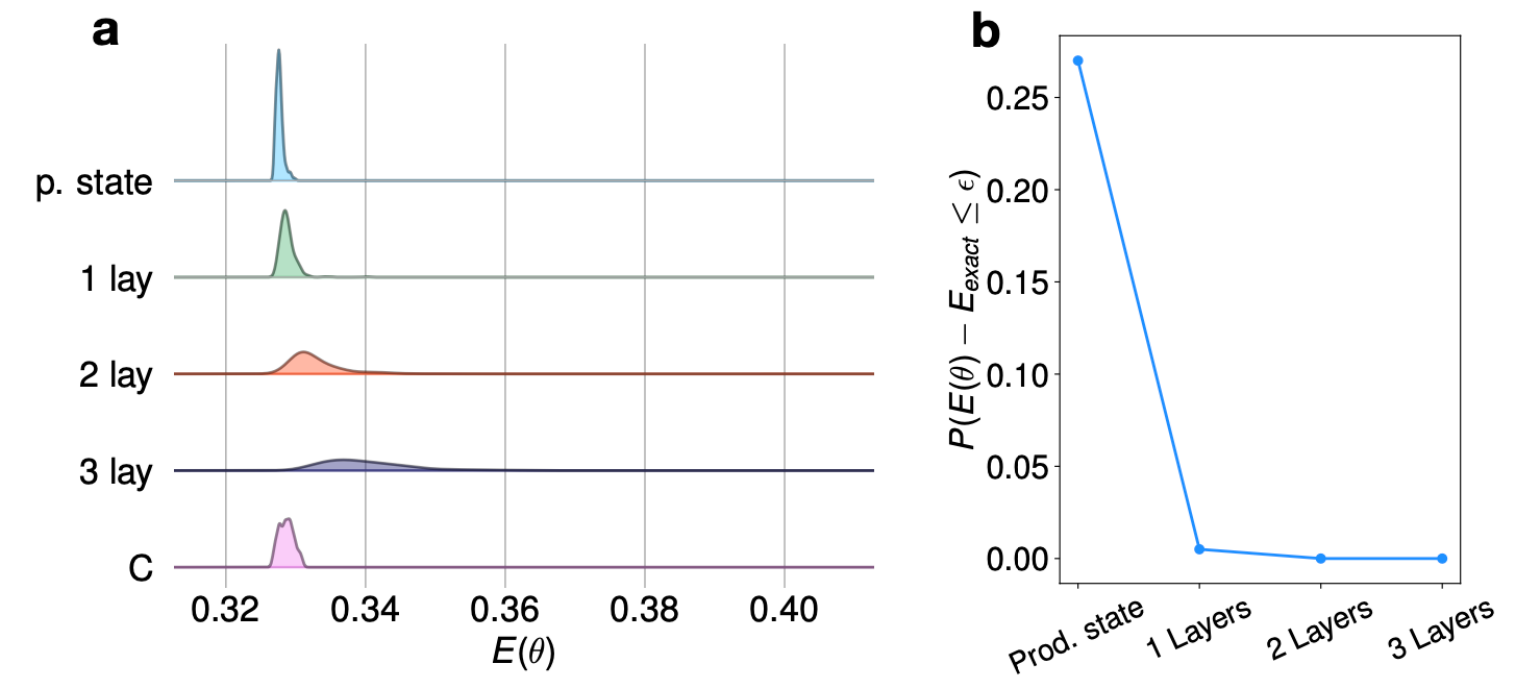} 
    \caption{\textbf{Gradient-free optimization for the reduced MNIST dataset.} (a) A KDE of $E(\boldsymbol{\theta})$ resulting from repeating the optimization procedure 300 times. We show the density of configurations with a cost $C(\bm{w})$. (b) A plot of the probability of success within a threshold $\epsilon=5\times10^{-4}$ demonstrates that the probability of finding the lowest $E(\theta)$ decreases with increasing circuit depth.}%
    \label{fig:gradfree}
\end{figure}

\begin{figure}%
   
    \includegraphics[width =\linewidth]{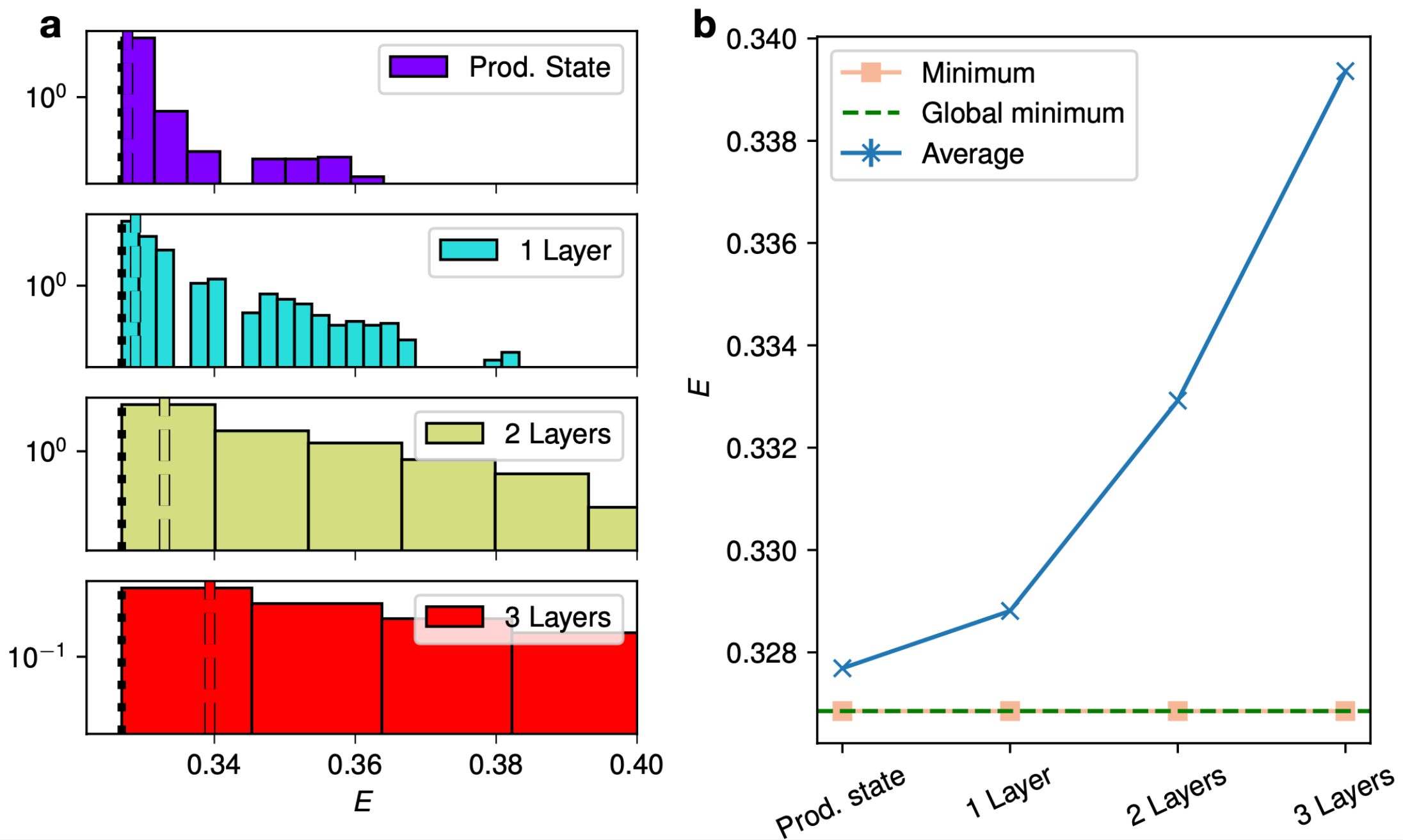} 
    \caption{
    \textbf{Sampling solutions optimized with gradient-free techniques for the reduced MNIST dataset.} (a) Histogram of the solutions found by the algorithm for different circuit depths. The mean energy $E(\theta)$ is shown as a vertical dashed line. The vertical dotted black lines depict the absolute minimum of the energy. (b) The average behaviour ($E(\theta)$) and minimum energy associated with the optimal BiNN as a function of circuit depth demonstrate that the algorithm finds the optimal solution with high probability. The error bars (smaller than symbols) represent one standard deviation. The averages, standard deviations, and minimum are taken over all the samples collected out of all the optimization runs,  i.e., over a total of $1000\times300$ samples.}%
    \label{fig:gradfreesamples}
\end{figure}

Here we investigate the effect of using gradient-free optimization on the performance of our algorithm on the reduced MNIST dataset problem. We consider a gradient-free optimizer based on a direct search based on probabilistic descent~\cite{grattonDirectSearchBased2015} as implemented within the BlackBoxOptim.jl package~\cite{Feldt2018}, which worked best among the gradient-free optimizers in the BlackBoxOptim.jl package. We perform 300 optimization runs for each of the circuits and explore the performance as a function of circuit depth. Each instance runs for 2000 iteration steps of the probabilistic descent algorithm, which requires evaluating the objective function ($E(\theta)$) for approximately $ 3500$ times independently of the depth of the circuit. This means that the complexity of one iteration step is significantly reduced with respect to a gradient-full calculation. The results are summarized in Fig.~\ref{fig:gradfree} and Fig.~\ref{fig:gradfreesamples}. 

Compared to results in Fig.~\ref{fig:MNIST}, we first mention that the optimization with the gradient-free method is less effective than gradient-based techniques. In addition, the average quality of the solutions found by the gradient-free method decreases significantly with increasing depth. This can be observed in Fig.~\ref{fig:gradfree}(a-b), where compared to Fig.~\ref{fig:MNIST}, the solutions are spread over higher values of energy for all circuit depths. Additionally, in Fig.~\ref{fig:gradfree}(b) it is evident that the probability of successfully finding low average $E(\theta)$ decreases sharply with increasing circuit depth.   

We also investigate the distribution of solutions contained in the final output states in Fig.~\ref{fig:gradfreesamples}(a-b). We note that unlike gradient-based simulations, where the probability distributions over configurations are sharply peaked near the qubit basis element $\sigma$ minimizing $E(\theta)$, our gradient-free simulations notably broaden the quantum state's probability distribution across computational basis states. This naturally results in an increase of the optimal $E(\theta)$. This effect becomes more pronounced with rising circuit depth, as depicted in Fig.~\ref{fig:gradfreesamples}(a). Here, a histogram of the energy (using all the samples derived from 300 optimizations each sampled 1000 times from their corresponding output state) illustrates an increasing broadening of the energy distribution with higher circuit depth. The average energies reported in Fig.~\ref{fig:gradfreesamples}(a-b) are computed over all the samples collected out of all the optimization runs,  i.e., over  a total $1000\times300$ samples.

Fortunately, despite the optimal $E(\theta)$ being notably higher than its gradient-full counterpart, our experiments consistently show that the optimal BiNN is reliably found in the final quantum state with high probability as observed in Fig.~\ref{fig:gradfreesamples}(b), where the minimum over all the samples coincide with the true minimum of the objective function $C$.


\section{Appendix: Noisy simulations under local depolarization} \label{ap:locdep}

Here we investigate the effect of noise on the performance of our algorithm on the reduced MNIST dataset problem. In our experiments, we apply a single-qubit depolarizing channel after the application of both single- and two-qubit gates on the qubits where the specific gate acts. The single-qubit depolarizing channel is given by
  \begin{align}
    \mathcal{\varrho} = \sum_{m=1}^M K_m \varrho K_m^\dag,
  \end{align}
  where $\{K_m\}$ is a set of Kraus operators with
  \begin{align}
    K_1 &= \sqrt{\frac{(1 - \lambda)}{4}}  \openone,\quad
    K_2 = \sqrt{\frac{\lambda}{ 3}} X\\
    K_3 &= \sqrt{\frac{\lambda}{ 3}} Y, \quad
    K_4 =  \sqrt{\frac{\lambda}{ 3}} Z.
  \end{align}
  Here, $\{X,Y,Z\}$ are the Pauli matrices and $\openone$ is the identity. The strength of the noise is given by $\lambda$. In our simulations we assume that the single- ($(\lambda_{\text{1-qubit}}$) and two-qubit ($(\lambda_{\text{2-qubit}}$) gate noise strengths are given by $(\lambda_{\text{1-qubit}},\lambda_{\text{2-qubit}} )=p\times(0.001, 0.00375)$, where $p$ is a positive parameter that re-scales the noise keeping the ratios of one and two qubit noise fixed.  Below, we either take $p=1$ and vary the circuit depth or fix the circuit depth vary $p$. We had assumed that the two-qubit depolarization is higher than the single-qubit one, in line with current experimental platforms.

\begin{figure}
\vspace{10pt} 
     \includegraphics[width =\linewidth]{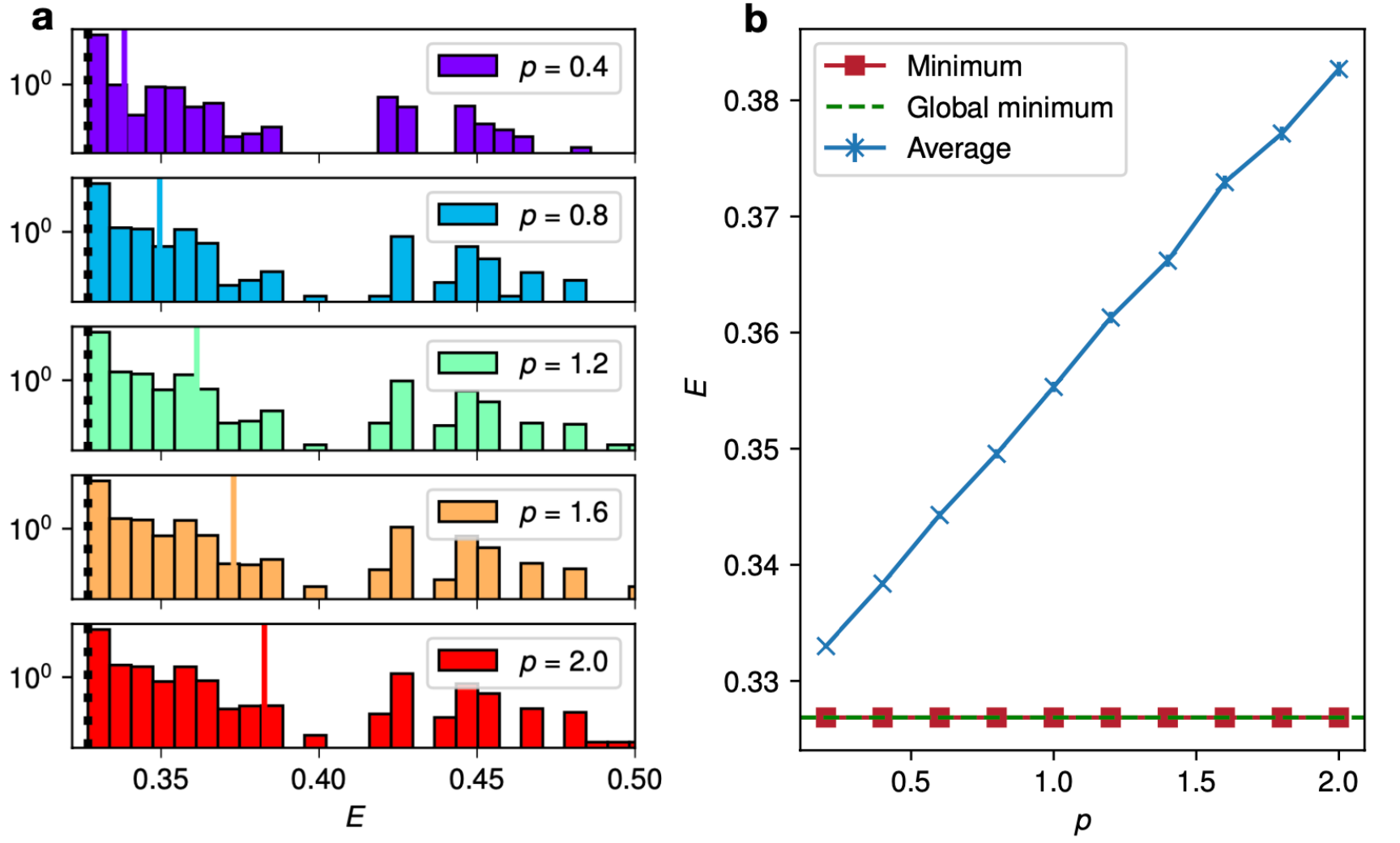} 
    \caption{\textbf{Robustness of noise-free optimized solutions to a local depolarization channel.} (a) Energy histogram of the solutions found by the algorithm for different values of noise amplitude $p$. The mean energy $E(\theta)$ is shown as a vertical full line. (b) $E(\theta)$ and histogram's minimum energy as a function of $p$ demonstrates that the algorithm finds the optimal solution with high probability despite the presence of noise. The error bars (smaller than symbols) represent one standard deviation. The averages, standard deviations, and minimum are taken over all the samples collected out of all the optimization runs,  i.e., over a total of $1000\times100$ samples.}%
    \label{fig:robustnesstonoise}
\end{figure}

First, we examine the robustness of solutions obtained from noise-free gradient-based simulations against local depolarization. We consider 100 instances of noise-free gradient-full optimization and introduce noise values $p=(0.2, 0.4, 0.6, 0.8, 1.0, 1.2, 1.4, 1.6, 1.8, 2.0)$. The circuit depth is set to 2, determined as optimal in Fig.~\ref{fig:MNIST}. Furthermore, each output state is sampled $N_s=1000$ times. In Fig.~\ref{fig:robustnesstonoise}(a), we display an energy histogram based on a total of $10^5$ samples collected from all $100$ circuit optimization runs. While noise-free simulations yield samples concentrated around a single optimal energy near the global minimum, depolarization causes solutions to spread across higher energies. This is seen to increase the average value of $E(\theta)$ linearly with noise amplitude $p$ (Fig.~\ref{fig:robustnesstonoise}(b)). In Fig.~\ref{fig:robustnesstonoise}(a), we also depict the absolute minimum of the energy (vertical dashed black line). Fortunately, we observe that despite the average energy $E(\theta)$ (vertical solid lines) rising with increasing $p$, the absolute minimum is still sampled with high probability within the noisy circuit for all $p$ values. Similar to our gradient-free experiment, this implies that the optimal solution remains accessible despite the presence of noise.

\begin{figure}%
     \includegraphics[width =\linewidth]{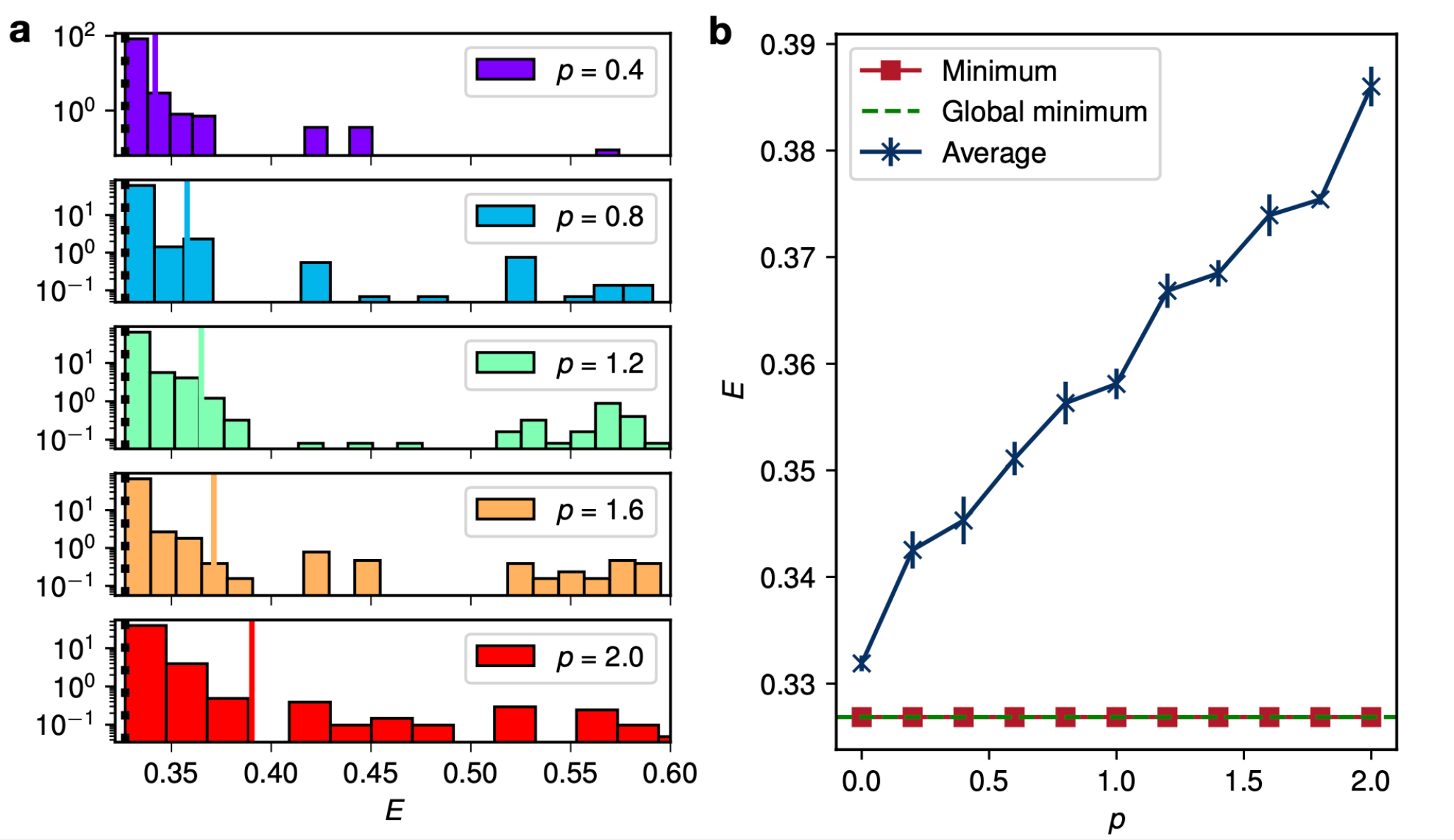} 
    \caption{\textbf{Noisy circuit optimized with gradient-free method} (a) Energy histogram of the solutions found by the algorithm for different values of noise amplitude $p$. The mean energy $E(\theta)$ is shown as a vertical dashed line. (b) $E(\theta)$ and histogram's minimum energy as a function of $p$  demonstrates that the algorithm finds the optimal solution with high probability despite the presence of noise. The error bars (smaller than symbols) represent one standard deviation. The averages, standard deviations, and minimum are taken over all the samples collected out of all the optimization runs,  i.e., over a total of $1000\times10$ samples.}%
    \label{fig:noiseoptimizationvsp}
\end{figure}

We also consider optimizing a noisy quantum circuit with the direct search based on probabilistic descent~\cite{grattonDirectSearchBased2015}. We carry out 10 optimization runs for each value of $p \in (0.0, 0.2, 0.4, 0.6, 0.8, 1.0, 1.2, 1.4, 1.6, 1.8, 2.0)$. Each instance runs for $2000$ iteration steps of the probabilistic descent algorithm, which requires evaluating the objective function ($E(\theta)$) for $\sim 3500$ times. Fixing the circuit depth to 2 and increasing $p$, we encounter a similar situation as in the gradient-free optimization of noiseless circuit example, namely that, despite the optimal $E(\theta)$ being notably higher than its gradient-full and noise-free counterpart, the optimal BiNN is reliably found in the optimized final quantum state with high probability. These results are summarized in Fig.~\ref{fig:noiseoptimizationvsp}.

\begin{figure}%
   
    \includegraphics[width =\linewidth]{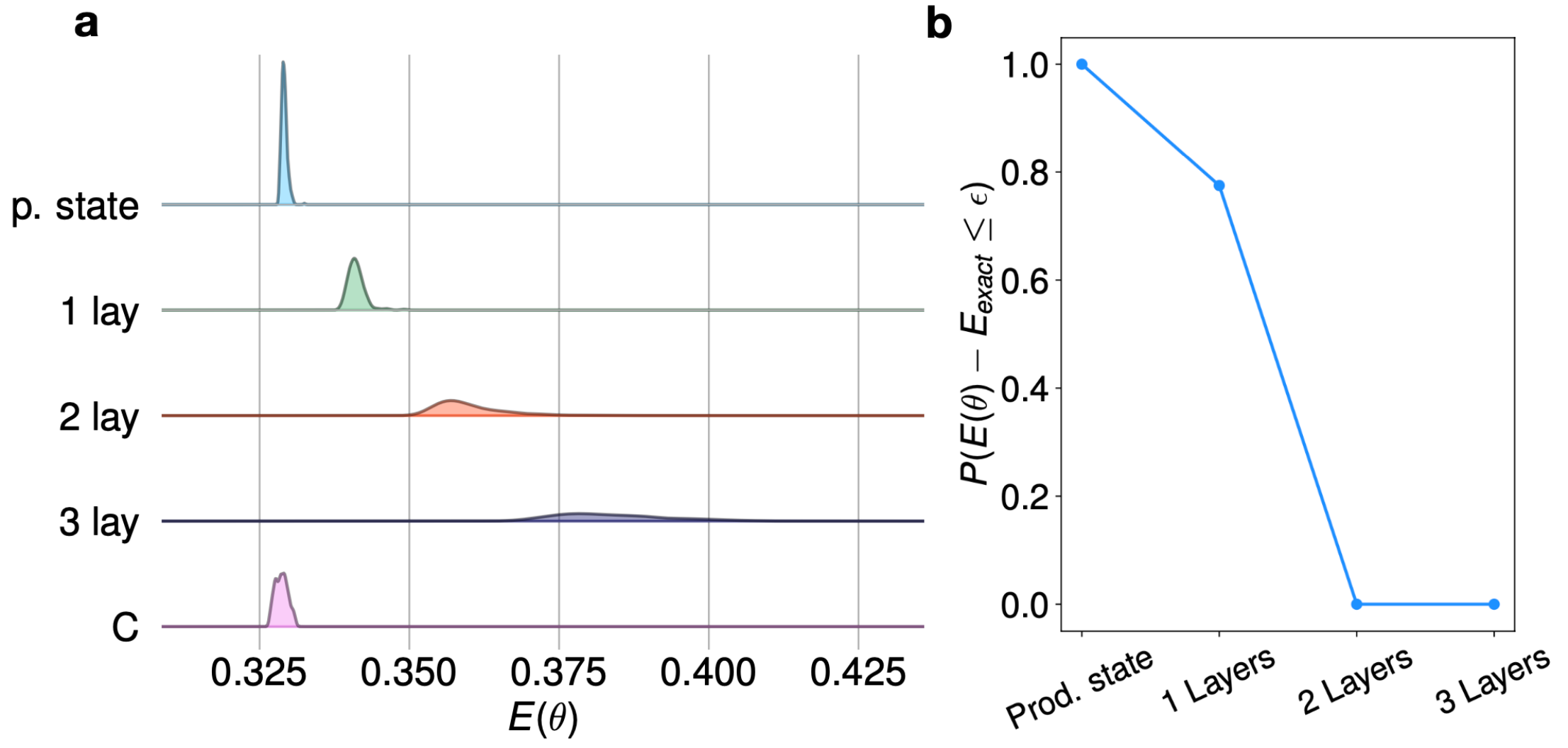} 
    \caption{\textbf{Gradient-free optimization of noisy circuits for the reduced MNIST dataset.} (a) A KDE of $E(\boldsymbol{\theta})$ resulting from repeating the optimization procedure 300 times. We also show the problem's exact density of configurations with a cost $C(\bm{w})$. (b) A plot of the probability of success within a threshold $\epsilon=1.5\times10^{-2}$ demonstrates that the probability of finding the lowest possible $E(\theta)$ decreases sharply with increasing circuit depth.}%
    \label{fig:gradfreenoisyKDE}
\end{figure}

\begin{figure}%
     \includegraphics[width =\linewidth]{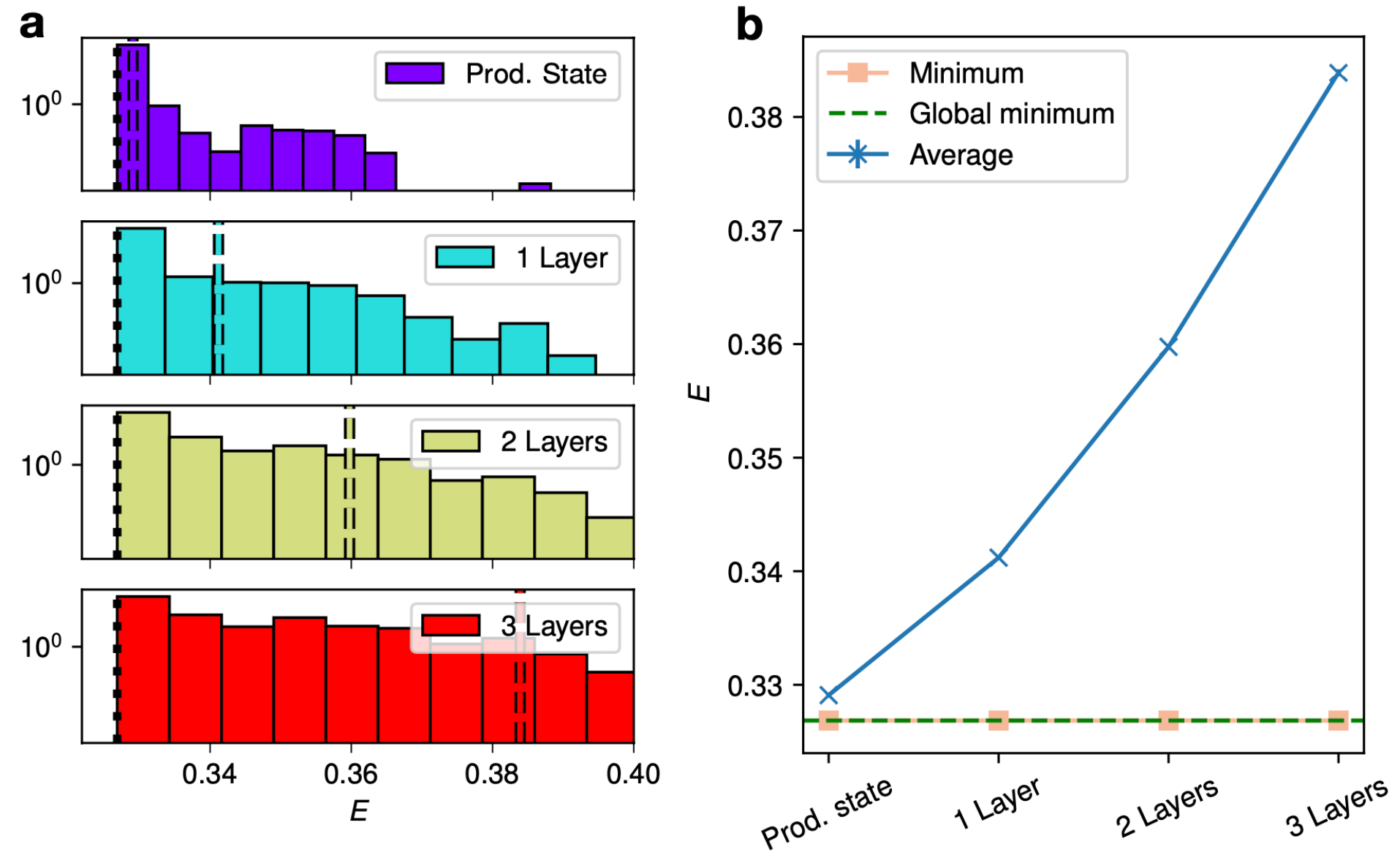} 
    \caption{\textbf{Noisy circuit optimized with gradient-free method as a function of depth} (a) Energy histogram of the solutions found by the algorithm for different circuit depths. The mean energy $E(\theta)$ is shown as a vertical dashed line. (b) $E(\theta)$ and histogram's minimum energy as a function of circuit depth demonstrates that the algorithm finds the optimal solution with high probability despite the presence of noise. The error bars (smaller than symbols) represent one standard deviation. The averages, standard deviations, and minimum are taken over all the samples collected out of all the optimization runs,  i.e., over a total of $1000\times300$ samples.}%
    \label{fig:noiseoptimizationvsdepth}
\end{figure}

Finally, we consider the noisy circuit optimization using a gradient-free based algorithm as a function of depth and fixed noise $p=1$. We perform $300$ optimization runs for each of the circuits and explore the performance as a function of circuit depth. As in our previous examples, each instance runs for 2000 iteration steps of the probabilistic descent algorithm, which requires evaluating the objective function ($E(\theta)$) for $\sim 3500$ times. The results are summarized in Fig.~\ref{fig:gradfreenoisyKDE} and Fig.~\ref{fig:noiseoptimizationvsdepth}.

In all of our experiments, we observe that the combination of noise and gradient-free optimization impacts the effectiveness of our approach when the depth of the circuit is increased. This is evidenced by the decreasing probability of finding $E(\theta)$ near the exact minimum (Fig.~\ref{fig:gradfreenoisyKDE}) increasing values of $E(\theta)$ (Fig.~\ref{fig:noiseoptimizationvsdepth}).  As done in our previous experiments, we also collect $1000$ samples from the final states of every optimization run and create an energy histogram (Fig.~\ref{fig:noiseoptimizationvsdepth}(a)). Despite the average energy $E(\theta)$ (vertical dashed lines in Fig.~\ref{fig:noiseoptimizationvsdepth}(a)) rising with increasing depth, the absolute minimum is still sampled with high probability within the noisy circuit for all circuit depths. This implies that the optimal solution remains accessible despite the presence of noise and lack of gradient information.

\section{Appendix: Computational time of the simulations} \label{ap:simt}

Here we briefly discuss the computational costs associated with our simulations. We consider the optimization of the MNIST example with selection of encoding (Fig.~\ref{fig:encoding}). This involves a quantum circuit with 19 qubits, which includes $4\times4$ weights, 1 bias, and 2 qubits for encoding selection. For a circuit with a depth of 6, a single optimization run with 500 iterations of the LBFGS algorithm on a MacBook Air Apple M2 processor with 24GB of RAM takes approximately 3.5 hours of walltime. On the Graham cluster from the Digital Research Alliance of Canada using a 4-core CPU with 24GB of RAM, the same simulations use 2.41 hours on average. In total, all of our circuit optimization experiments consumed approximately 3.44 CPU core years.

Additionally, we examine the computational cost of full enumeration, which we utilize for constructing the operator $\hat{C}$ and obtaining the exact solution. This calculation scales exponentially with the number of binary variables in the problem. For instance, the MNIST problem with 19 qubits takes approximately 3 minutes, making it computationally faster than the classical simulations of the variational algorithm at this problem size.

However, it's important to note that for larger problems, the cost of running the variational algorithm (either on a real quantum device for circuits of polynomial depth or classically for shallow quantum circuits) will not increase exponentially if we restrict the maximum number of optimization iterations, e.g., to a constant. This is because for large problems, we would require the evaluation of the cost function over a small number of projective measurements, instead of its exact computation, which requires constructing the operator $\hat{C}$, an operation that is exponential in the number of variables.

Therefore, there exists a certain system size beyond which it becomes computationally less expensive to run a variational algorithm than to perform full enumeration. Finally, while the variational algorithm provides no performance guarantees in our setting and is considered a heuristic algorithm, we can control its execution time. The results can be assessed for overfitting using a test dataset, as demonstrated in Fig.~\ref{fig:overfitting}.

\end{document}